\documentclass{emulateapj}

\newcommand{\boo}{Bo\"otes}

\newcommand{\ltsim}{\raisebox{-0.6ex}{$\,\stackrel
        {\raisebox{-.2ex}{$\textstyle <$}}{\sim}\,$}}
\newcommand{\gtsim}{\raisebox{-0.6ex}{$\,\stackrel
        {\raisebox{-.2ex}{$\textstyle >$}}{\sim}\,$}}
\newcommand{\simlt}{\raisebox{-0.6ex}{$\,\stackrel
        {\raisebox{-.2ex}{$\textstyle <$}}{\sim}\,$}}
\newcommand{\simgt}{\raisebox{-0.6ex}{$\,\stackrel
        {\raisebox{-.2ex}{$\textstyle >$}}{\sim}\,$}}

\newcommand{\Msun}{M_\odot}
\newcommand{\Lsun}{L_\odot}

\newcommand{\Mdot}{\dot M}

\newcommand{\ntot}{2603}

\newcommand{\nfaint}{703}

\newcommand{\nz}{86} 

\slugcomment{Draft 1 Version 2007 Sep 7}
\slugcomment{Submitted Version: 2007 Nov 6}
\slugcomment{Revised Version: 2008 Jan 10}
\slugcomment{Accepted for Publication in ApJ: 2008 Jan 10}

\shorttitle{$z\approx 2$ DOGs}
\shortauthors{Dey et al.}

\begin{document}

\title {A Significant Population of Very Luminous Dust-Obscured Galaxies at Redshift $z\sim 2$}

\author{Arjun Dey\altaffilmark{1}, 
B. T. Soifer\altaffilmark{2,3}, 
Vandana Desai\altaffilmark{2},
Kate Brand\altaffilmark{4},
Emeric LeFloc'h\altaffilmark{5,6},
Michael J. I. Brown\altaffilmark{7},
Buell T.\ Jannuzi\altaffilmark{1}, 
Lee Armus\altaffilmark{3},
Shane Bussmann\altaffilmark{8},
Mark Brodwin\altaffilmark{1},
Chao Bian\altaffilmark{2},
Peter Eisenhardt\altaffilmark{9},
Sarah J. Higdon\altaffilmark{10},
Daniel Weedman\altaffilmark{11},
S. P. Willner\altaffilmark{12}
}

\altaffiltext{1}{National Optical Astronomy Observatory, 950 N. Cherry Ave., Tucson, AZ 85719; dey@noao.edu}
\altaffiltext{2}{Caltech Optical Observatories, California Institute of Technology, Pasadena, CA 91125}
\altaffiltext{3}{Spitzer Science Center, California Institute of Technology, MS 220-6, Pasadena, CA 91125}
\altaffiltext{4}{Space Telescope Science Institute, Baltimore, MD 21218; Giacconi Fellow}
\altaffiltext{5}{Institute for Astronomy, University of Hawaii, Honolulu, HI 96822}
\altaffiltext{6}{Spitzer Fellow}
\altaffiltext{7}{School of Physics, Monash University, Clayton, Victoria 3800, Australia}
\altaffiltext{8}{Steward Observatory, University of Arizona, Tucson, AZ 85721}
\altaffiltext{9}{Jet Propulsion Laboratory, California Institute of Technology, MC 169-327, 4800 Oak Grove Drive, Pasadena, CA 91109}
\altaffiltext{10}{Georgia Southern University, P.O.~Box 8031, Statesboro, GA}
\altaffiltext{11}{Astronomy Department, Cornell University, Ithaca, NY 14853}
\altaffiltext{12}{Harvard-Smithsonian Center for Astrophysics, 60 Garden Street, Cambridge, MA 02138}

\begin{abstract}

Observations with the {\it Spitzer Space Telescope} have recently revealed a significant population of high-redshift ($z\sim2$) dust-obscured galaxies (DOGs) with large (rest-frame) mid-infrared to ultraviolet luminosity ratios. Due to their optical faintness, these galaxies have been previously missed in traditional optical studies of the distant universe.  We present a simple method for selecting this high-redshift population based solely on the ratio of the observed mid-infrared 24$\mu$m to optical $R$-band flux density. We apply this method to observations of the $\approx 8.6~{\rm deg}^2 $ Bo\"otes Field of the NOAO Deep Wide-Field Survey, and uncover $\approx$2,600 DOG candidates (i.e., a surface density of 0.089~arcmin$^{-2}$) with 24$\mu$m flux densities $F_{\rm 24\mu m} \ge 0.3$mJy and $(R-[24])\ge 14$ (i.e., $F_\nu({\rm 24\mu m})/F_\nu(R) \simgt 1000$). These galaxies have no counterparts in the local universe. They become a larger fraction of the population at fainter 24$\mu$m flux densities, increasing from 7$\pm$0.6\% of sources at $F_{24\rm \mu m}\ge 1$~mJy to $\approx 13\pm1$\% of the population at $\approx$ 0.3~mJy.  These galaxies exhibit evidence of both star-formation and AGN activity, with the brighter 24$\mu$m sources being more AGN-dominated. Their mid-infrared spectral energy distributions range from power-laws (likely AGN-dominated at mid-IR wavelengths) to systems showing a ``bump'', the latter likely resulting from the redshifted 1.6$\mu$m peak characteristic of most stellar populations. Using primarily the W. M. Keck Observatory and {\it Spitzer}, we have obtained spectroscopic redshifts for \nz\ objects within
this sample, and find a broad redshift distribution which can be modeled as a Gaussian centered at $\bar{z}\approx1.99\pm0.05$ and $\sigma(z)\approx0.45\pm0.05$. The space density of this population is $\Sigma_{\rm DOG} (F_{\rm 24\mu m}\ge 0.3~{\rm mJy}) = (2.82\pm 0.05)\times 10^{-5} h_{70}^3~ {\rm Mpc^{-3}}$, similar to that of bright sub-millimeter-selected or UV-selected galaxies at comparable redshifts. These redshifts also imply very large luminosities, with a sample median $\nu L_\nu(8\mu{\rm m})\approx 4\times 10^{11}\Lsun$, implying ${\rm 8\mu m - 1mm}$ luminosities of $L_{\rm IR}\gtsim 10^{12-14}\Lsun$ for the population. The infrared luminosity density contributed by this relatively rare DOG population is ${\rm log}(L_{\rm IR})\approx 8.23^{+0.18}_{-0.30}$. This is $\approx60^{+40}_{-15}$\% of that contributed by $z\sim2$ ultraluminous infrared galaxies (ULIRGs, with $L_{\rm IR}>10^{12}\Lsun$), and suggests that our simple selection criterion effectively identifies a significant fraction of $z\sim 2$ ULIRGs. This IRLD is also $\approx26\pm14$\% of the total contributed by all $z\sim2$ galaxies, and comparable to that contributed by the luminous UV-bright star-forming galaxy populations at $z\approx 2$. We suggest that these DOGs are the progenitors of luminous ($\sim 4L^*$) present-day galaxies and are undergoing an extremely luminous, short-lived phase of both bulge and black hole growth. They may represent a brief evolutionary phase between sub-millimeter-selected galaxies and less obscured quasars or galaxies.  

\end{abstract}

\keywords{galaxies:formation--galaxies:high-redshift--galaxies:starburst}

\section{Introduction}

Infrared-luminous sources (with $L_{\rm IR}>10^{11}{\rm L_\odot}$) dominate the bright end of the bolometric luminosity function of galaxies in the local universe \citep[]{soi1986}.  Their role in the story of galaxy evolution is not yet understood, but it is likely that their prodigious luminosities are the product of an extremely active phase during which these systems form stars and/or grow their central black holes at a rapid rate \citep[]{san1988}.  Did all large galaxies exhibit such a phase in their early formative stages, or is this active phase characteristic of only some of the most massive systems?  To address these questions, we need to find and study the primary high-redshift galaxy populations contributing to the mid-IR and far-IR counts and understand their relation to the local galaxy population.

We know a great deal about the local population of IR-luminous galaxies, due in large part to the pioneering observations made with the {\it Infrared Astronomical Satellite} \citep[{\it IRAS}; e.g.,][]{soi1987a,san1988,san1996}. Their mid-IR number counts show strong evolution \citep[e.g.,][]{elb1999,chary2001,pap2004,chary2004,marl2004}. Although rare locally, infrared luminous galaxies become an increasing fraction of the galaxy population at high redshift, dominating the IR energy density at redshifts $z\sim1$ \citep[e.g.,][]{fra2001,lag2004,gru2005,lef2005,per2005}. 

Far less is known about their higher redshift counterparts despite several ground-breaking studies.  Prior to the launch of the {\it Spitzer Space Telescope}, our knowledge of high-$z$ IR-luminous galaxies was derived from studies of a handful of the most extreme {\it IRAS} sources, constraints on the redshift evolution out to $z\sim 1$ \citep[largely due to {\it ISO}; e.g.,][]{fra2001,chary2001,lag2003}, small samples of galaxies discovered at sub-millimeter wavelengths \citep[e.g.,][]{sma2002,bla2002,cha2004b,cha2005}, and several galaxies discovered as a result of their extreme optical-to-near-infrared colors \citep[the dusty `Extremely Red Objects', or EROs, with $R-K>6$; e.g.,][]{hu1994,dey1999,mcc2004a}. 

The launch of the {\it Spitzer Space Telescope} has revolutionized studies of the dusty galaxy population. It has provided the first mid-IR spectroscopy of dust emission in high-redshift star-forming galaxies and AGN, and revealed populations of galaxies that are extremely faint at (observed) visible wavelengths \citep[e.g.,][Higdon et al. and Desai et al., both in preparation]{hou2005,yan2005,yan2007}. Many of these galaxies lie at redshifts $z\simgt 1$, and their mid-IR spectra reveal either strong silicate absorption features, suggestive of an AGN, or emission from polycyclic aromatic hydrocarbons (PAHs) which generally arise in photodissociation regions associated with star-forming molecular clouds. A significant fraction of these active high-redshift galaxies are under-luminous at rest-frame UV wavelengths, suggesting that the prodigious star-formation and / or AGN accretion luminosity is hidden by dust that reprocesses and radiates the bulk of the energy at far-IR wavelengths \citep[e.g.,][]{san1996}. 

In this paper, we present a simple and economical method for selecting these high-redshift dusty populations, based solely on the ratio of the observed 24$\mu$m to optical flux ratio. This selection, applied to a wide area survey, results in a large, significant population of extremely luminous high-redshift galaxies for which there are no local counterparts. These are systems with unusually red spectral energy distributions, which we interpret as being the result of extinction by large columns of dust. These galaxies appear to be largely absent from (rest-frame) UV-selected high-redshift galaxy samples. We have measured a large number of spectroscopic redshifts for this population, and confirm that they lie at redshifts $z\sim2$. Given their large mid-infrared luminosities and implied far-infrared luminosities, these dust obscured galaxies may be undergoing an intense burst of star formation, accretion activity (onto a nuclear black hole) or plausibly both. We speculate that this population of objects, which may be selected simply on the basis of optical to mid-infrared flux ratio, are good candidates for systems in the throes of formation of their stellar spheroids and nuclear black holes. Despite their rarity, this population contributes a significant fraction of the infrared luminosity density at $z\sim 2$.

This paper is structured as follows. In \S2, we briefly describe the observations from which the sample is drawn and discuss the selection criteria we use to isolate the candidate high-redshift dust-obscured galaxy (hereinafter DOG) population. In \S3, we present the main results: the optical-to-mid-infrared color distribution of the DOGs, their spectral energy distributions, redshift distribution and rest-frame UV and IR luminosities. In \S4, we discuss the nature of the DOG population, derive their space density and their contribution to the $z\approx 2$ infrared luminosity density, and discuss their possible relationship to other $z\approx 2$ galaxy populations. We summarize our findings in \S5. 

Throughout this paper we adopt a cosmology with ${\rm H_0 = 70~km\ s^{-1}\ Mpc^{-1}}$, $\Omega_m = 0.3$, $\Omega_\lambda = 0.7$. Magnitudes quoted in this paper are in Vega units, except where explicitly defined otherwise (e.g., in the discussion of rest-frame 2200\AA\ luminosities).

\section{Observational Data and Sample Selection}

\subsection{Imaging Data}

The study described in this paper focusses on a sample drawn from the mid-infrared and optical observations of the \boo\ Field of the NOAO Deep Wide-Field Survey \citep[NDWFS\footnote{See also http://www.noao.edu/noaodeep/}; ][Jannuzi et al., in prep., Dey et al., in prep.]{ndwfs}. The NDWFS is a deep, ground-based, optical and near-infrared imaging survey of two 9.3 square degree fields, one in \boo\ and one in Cetus, conducted using the 4m and 2.1m telescopes of the National Optical Astronomy Observatory. The survey reaches median 5$\sigma$ point-source depths in the $B_W, R, I$ and $K$ bands of $\approx$27.1, 26.1, 25.4 and 19.0 mag (Vega) respectively (see the NOAO Science Archive at {\tt http://http://archive.noao.edu/nsa/} for details). The data products for the NDWFS will be described elsewhere (Dey et al., in prep.; Jannuzi et al., in prep). 
The NDWFS astrometry is tied to the reference frame defined by stars from the USNO A-2 catalog \citep[]{usnoA2}. 

{\it Spitzer} has mapped 8.61~deg$^2$ of the NDWFS \boo\ field at 24, 70 and 160\micron\ using the Multiband Imaging Photometer for Spitzer  \citep[MIPS; ][]{rie2004}  to 1$\sigma$ rms depths of 0.051mJy, 5mJy and 18mJy respectively. The data were reduced by the MIPS GTO team. Details of the survey, including the mapping strategy, data reduction and the resulting catalog, will be discussed elsewhere. The entire \boo\ field was also mapped at 3.6, 4.5, 5.8 and 8.0\micron\ using the Infrared Array Camera \citep[IRAC; ][]{faz2004} on {\it Spitzer}; details of the IRAC observations may be found in \citet[]{eis2004}. Both the MIPS and IRAC data are astrometrically calibrated to the 2MASS astrometric frame. There are roughly 22,000 24$\mu$m sources in the \boo\ field down to the 80\% completeness limit of 0.3~mJy. 

The 24$\mu$m source catalogs were matched to the optical catalogs (after accounting for a small but significant relative offset of $\Delta\alpha=0{\farcs}38$ and $\Delta\delta=0\farcs14$ between the two frames). Optical photometry was determined for each 24$\mu$m source using the images of the third data release (DR3) of the NDWFS. In order to measure accurate optical colors in small aperture photometry, we smoothed the optical NDWFS images so the delivered point spread function across the entire Bo\"otes field was a $1.35^{\prime\prime}$ FWHM Moffat profile with a $\beta$ parameter of $2.5$. Using our own code, we measured $4^{\prime\prime}$ diameter aperture photometry at the $24~\micron$ position of each source. The $R$-band magnitudes of the brighter sources, i.e. with $R\le 25$, are the `AUTO' magnitudes derived by SourceExtractor \citep[]{ber1996}. For fainter sources and for mid-IR sources without optical counterparts in the optical catalogs, photometric limits were measured directly from the optical stacked images in 4\arcsec\ diameter apertures. In cases where the 4\arcsec\ diameter aperture magnitude (or limit) was indeterminate or greater than the 50\% completeness limit of our optical data, we replaced the estimated magnitude with the brighter of the locally measured 3$\sigma$ sky noise or the 50\% completeness limit.  The background and uncertainties of the photometry were determined by measuring the sigma-clipped mean and RMS of fluxes measured in sixteen $4^{\prime\prime}$ apertures located in a $45^{\prime\prime}$ radius circle around each object. 

\subsection{Sample Selection}

\begin{figure}[t]
\epsscale{0.8}
\plotone{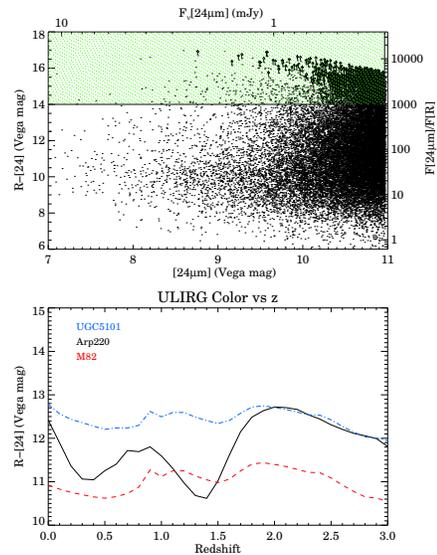}
\caption{{\it Top panel:} ($R$-[24]) color vs. 24$\mu$m magnitude distribution for all sources in the NDWFS Bo\"otes Field. The bottom and top abscissae show the 24$\mu$m magnitude and flux density respectively, and the left and right ordinates show the color in magnitudes and the $F_{\rm 24\mu m}/F_R$ flux density ratio respectively. The objects with ($R$-[24])$\ge$14 lie above the horizontal line. {\it Bottom panel:} The expected variation of color with redshift of two local ULIRGs and the nearby starburst galaxy M82. ($R$-[24])$\ge$14 sources have no local counterparts.
\label{color}}
\end{figure}

\begin{figure}[t]
\epsscale{1.0}
\plotone{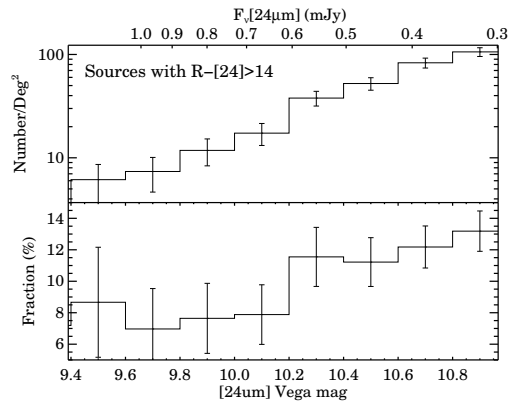}
\caption{The differential sky density (top panel) and fraction (bottom panel) of 24$\mu$m sources with colors $(R-[24])\ge14$ mag as a function of 24$\mu$m flux density. The fraction rises steadily as a function of decreasing 24$\mu$m flux density beyond $F_{\rm 24\mu m}\approx 1$mJy.
\label{redfraction}}
\end{figure}

Our sample is selected from the 8.140~deg$^2$ region of the Bo\"otes field for which there exists both good {\it Spitzer}/MIPS 24$\mu$m and KPNO $R$-band imaging. Regions around bright, saturated stars were masked out. $(R-[24])$ colors (in the Vega system)\footnote{We define the 24$\mu$m Vega magnitude as $[24]\equiv -2.5~{\rm log_{10}}\left(F_{\rm 24\mu m}/7.29{\rm Jy}\right)$. $F_{\rm 24\mu m}=1$mJy (0.3mJy) corresponds to a Vega magnitude of 9.66 (10.96).} computed for all 24$\mu$m sources  are shown as a function of 24$\mu$m flux density ($F_{24\rm \mu m}$) in Figure~\ref{color}. The points appear to spread toward redder colors at fainter $F_{24\rm \mu m}$. In order to quantify this effect, we define a fiducial color of $(R-[24]) = 14$ corresponding to $F_{\rm 24\mu m}/F_R\ge 982$, chosen because this is redder than the color of most ultra-luminous infrared galaxies (ULIRGs; $L_{\rm IR}\ge 10^{12}\Lsun$) at all redshifts (Figure~1b). We measure the (differential) fraction of the source population redder than this limit as a function of $F_{24\rm \mu m}$ and find that the fraction increases to fainter flux densities, rising from 7$\pm$0.6\% of the population for all sources with $F_{24\rm \mu m} \ge 1$mJy to 13$\pm$1\% at $F_{24\rm \mu m} \approx 0.3$mJy (Figure~\ref{redfraction}).  

We therefore selected our primary sample of dust-obscured galaxy candidates from the Bo\"otes Field of the NOAO Deep Wide-Field survey based solely on the following two criteria:
\begin{eqnarray}
 F_{\rm 24\mu m}\ge 0.3~{\rm mJy} \\
(R-[24]) \ge 14 {\rm (Vega)~mag}
\end{eqnarray}
There are \ntot\ sources satisfying these criteria in the \boo\ field. Of these, \nfaint\ are fainter than the 50\% completeness limit of our optical imaging, and the $(R-[24])$ values are therefore lower limits. We refer to this set of 2603 galaxies as our ``total sample'', and use it for estimating the infrared luminosity density contributed by this population. This large number of sources corresponds to an average surface density of $\approx 0.089$ sources arcmin$^{-2}$. 

As mentioned above, a subset of the Bo\"otes Field has been mapped by IRAC \citep[]{eis2004}. The total area of the combined $B_W$, $R$, $I$, IRAC 3.6$\mu$m and 4.5$\mu$m, and MIPS 24$\mu$m observations is (after additionally masking out bright stars and bad regions) 6.76~deg$^2$. All of the galaxies selected by criteria (1) and (2) above that lie within this region are detected in at least the 3.6$\mu$m and 4.5$\mu$m IRAC bands. We refer to this resulting sample of 1,882 sources as the ``photometric sample''. This latter sample is used below in the discussion of the IRAC color-color diagrams and for constructing robust spectral energy distributions. 

\subsection{Spectroscopic Data}

Members of this extremely red 24$\mu$m source population were targeted for spectroscopic observations using the Infrared Spectrometer on {\it Spitzer} \citep[IRS;][]{irs}, the Low Resolution Imaging Spectrometer  \citep[LRIS;][]{lris}, the Deep Imaging Multi-Object Spectrograph \citep[DEIMOS;][]{deimos}, and the Near Infrared Spectrometer \citep[NIRSPEC;][]{nirspec} on the telescopes of the W.~M.~Keck Observatory, and the Near-Infrared Imaging Spectrograph (NIRI) on the Gemini-N telescope.  The \boo\ field is also the target of the AGN and Galaxy Evolution Survey (AGES; Kochanek et al. 2008 in preparation), which has measured over 20,000 redshifts for optically bright galaxies ($I\simlt 20$) and AGN ($I\simlt 21.5$) using Hectospec \citep[]{hectospec} on the MMT. The IRS observations focused primarily on the mid-IR bright sources which were undetected at optical wavelengths, targeting objects with $F_{\rm 24\mu m}\gtsim 0.75$~mJy; preliminary results have been presented by \citet{hou2005} and \citet{wee2006}. Redshifts were measured from the mid-IR spectra by fitting either the strong (rest-frame) 9.7$\mu$m silicate absorption feature, or the 7.7$\mu$m PAH emission features. Typical uncertainties for the IRS redshifts are $\Delta z \simlt \pm0.2$. We refer the reader to the \citet{hou2005,wee2006} and Desai et al.~(2008, in prep) papers for details. 

The Keck optical observations targeted members of the $(R-[24])\ge14$ population as well as randomly selected  24$\mu$m sources down to the 24$\mu$m flux density limit of 0.3~mJy. The details of the Keck spectroscopic observations will be presented elsewhere (Soifer et al. 2007, in preparation) and the resulting redshift distribution of the overall $F_{\rm 24\mu m}\ge 0.3$~mJy source population is presented in Desai et al. (2008, ApJ, submitted). Redshifts of the sample discussed in this paper were derived from the optical spectroscopy by (typically) identifying either the [OII]$\lambda\lambda$3726,3729 doublet in the far red (for the $z\simlt 1.6$ galaxies) or the Ly$\alpha$ line in the blue for the $z\simgt 1.8$ objects. The near-infrared spectroscopic redshifts are based on the detections of the H$\alpha$+[NII] or [OIII] emission lines \citep[see][for examples]{bra2007}. Typical uncertainties for the optical and near-IR measured redshifts are $\Delta z \simlt 0.002$. 

Based on these various observations (but primarily on the {\it Spitzer} and Keck efforts), we have now measured a total of \nz\ spectroscopic redshifts for the sample of candidate dust-obscured galaxies (i.e., $F_{\rm 24\mu m}\ge 0.3$~mJy and $(R-[24])\ge14$) in the Bo\"otes field. Roughly half the redshifts (41/\nz) are the result of {\it Spitzer}/IRS observations, and half come from ground-based optical (38) or near-IR (7) spectroscopy.  There are only 5 galaxies which have redshift measurements from both {\it Spitzer}/IRS and Keck. These 5 have a mean redshift difference of $\Delta z\equiv z_{\rm Keck}-z_{\rm IRS}\approx -0.04$ (with an uncertainty on the mean of $\pm0.15$) and a standard deviation of $\sigma_{\Delta z}\approx 0.3$. Hence, we do not detect any statistically significant systematic redshift offset for the IRS redshifts, and adopt them as the true redshifts of the galaxies for the purposes of this paper.

We also compared the spectroscopically measured redshifts with photometric redshift estimates from various template fitting algorithms, and found a very large scatter between photometric redshift estimates and the spectroscopic measurements; photometric redshifts are not reliable for this extreme population. Given the uncertainties inherent in fitting templates to spectral energy distributions dominated by power laws and only weak features, we do not attempt to estimate photometric redshifts \citep[cf.][]{per2005,pol2006}, and instead use the spectroscopically measured redshift distribution directly.

\begin{figure}[t]
\epsscale{1.0}
\plotone{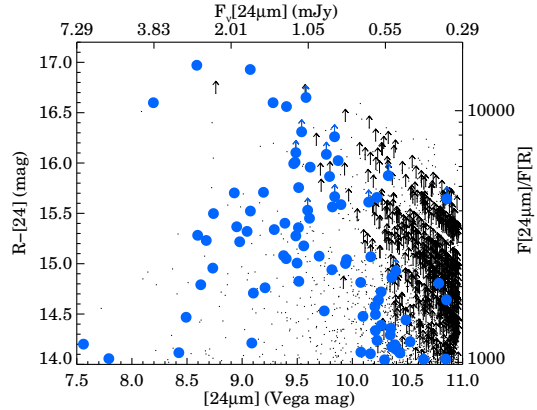}
\caption{Same as figure~\ref{color}, but restricted to sources with colors $(R-[24])\ge 14$. The dots represent the entire sample and the solid (blue) circles represent the galaxies with measured spectroscopic redshifts. The upward arrows show sources with upper limits on their $R$-band magnitudes.  The spectroscopic redshifts sample the entire range of color and magnitude. 
\label{zspecDOGs}}
\end{figure}

The galaxies for which we have measured redshifts sample the full range of $F_{\rm 24\mu m}$ and  $(R-[24])$ color (Figure~\ref{zspecDOGs}). For the objects with measured $(R-[24])$ colors (i.e., excluding the galaxies with only upper limits), the color distribution of galaxies with redshifts is a reasonable sampling of the full color distribution of the sample (i.e., a two-sided Kolmogorov-Smirnov test finds that the color distributions are similar at the 38\% level). The sampling of the 24$\mu$m flux density distribution is much poorer. This is because a significant fraction of the source redshifts were measured with {\it Spitzer}/IRS, and thus are restricted to the galaxies with $F_{\rm 24\mu m}\simgt 0.5$~mJy. Nevertheless, the spectroscopic sampling does span the full range in both color and 24$\mu$m flux density, and we assume for the purposes of this paper that the resulting redshift distribution is a fair representation of the overall population.



\section{Results}

In this section we discuss the colors and overall spectral energy distributions (hereafter SEDs) of the $R-[24]\ge 14$ galaxy population. We also present the redshift distribution, and under the assumption that the measured distribution applies to the entire population, we estimate the rest-frame UV, mid-IR and IR luminosity distributions. Finally, we discuss the X-ray emission from the sample, with the aim of understanding whether AGN play a role in the energetics.

\subsection{Colors and SEDs}

\subsubsection{Mid-Infrared Colors}

\begin{figure}[t]
\epsscale{0.65}
\plotone{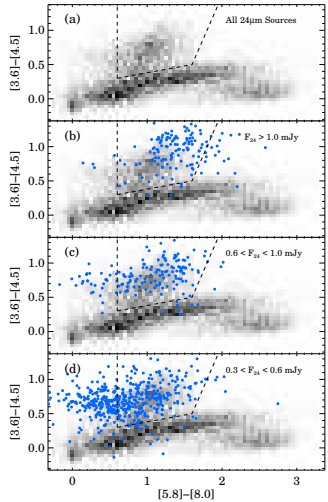}
\caption{{\it (a)} The mid-infrared color distribution of all 24$\mu$m sources in the NDWFS Bo\"otes field with flux densities $F_{\rm 24\mu m}\ge 0.3$~mJy, shown as a grey scale. This greyscale is repeated in all panels. The dashed lines demarcate the region occupied by bright AGN as defined by \citet[]{ste2005}. {(b)} The mid-infrared color distribution of the DOGs ($(R-[24])\ge14$ sources; shown as solid circles) with 24$\mu$m flux densities $F_{\rm 24\mu m}\ge 1$~mJy. {\it (c)} Same as {\it (b)}, but for DOGs with $0.6\le F_{\rm 24\mu m} < 1$~mJy. {\it (d)} Same as {\it (b)}, but for DOGs with $0.3\le F_{\rm 24\mu m} < 0.6$~mJy. The DOGs span a large range of [5.8]-[8.0] color, and the color is a function of the observed 24$\mu$m flux density, with the fainter sources having bluer colors. Notice that there is a continuum of sources across the ``AGN boundary'' of \citet[]{ste2005} at ([5.8]-[8.0])=0.6.
\label{iraccolor}}
\end{figure}

Several recent studies have demonstrated that the mid-infrared colors of galaxies can provide an effective discriminant between `normal' galaxies and unobscured AGN \citep[e.g.,][]{lac2004,ste2005}.  
The top panel of Figure~\ref{iraccolor} shows the distribution of mid-infrared color for all the 24$\mu$m sources with $F_{\rm 24\mu m}\ge 0.3$~mJy in the Bo\"otes NDWFS field. The color distribution is similar to that seen by \citet[]{ste2005}, showing a sequence of low-redshift star-forming galaxies at `blue' [3.6]$-$[4.5] colors, and a second sequence at redder colors that largely overlaps the `AGN wedge' \citep[shown by the dashed line; see][]{ste2005}. 

The mid-infrared colors of the sources with $(R-[24])\ge 14$ are shown in the lower panels of figure~\ref{iraccolor}. The $(R-[24])\ge 14$ population lies at redder [3.6]$-$[4.5] colors (with the majority of objects having [3.6]$-$[4.5]$>$0.5), suggesting that this population lies at high redshift ($z\simgt 1$) or are AGN \citep[]{ste2005}. Their positions in the color-color diagram also overlap the `AGN wedge', but spill across the boundary of the wedge toward bluer [5.8]$-$[8.0] colors. The figure also demonstrates that the location of the population in the color-color diagram is a function of $F_{\rm 24\mu m}$, with the apparently brighter sources (with $F_{\rm 24\mu m}\simgt 1$mJy) lying toward redder [5.8]$-$[8.0] colors, and the fainter sources (with $F_{\rm 24 \mu m}\simlt 0.6$mJy) having bluer [5.8]$-$[8.0] colors. The sequence appears to be continuous across the boundary of the `AGN wedge' at ([5.8]$-$[8.0])=0.6, with no obvious break in the population. The median values of the ([3.6]$-$[4.5],[5.8]$-$[8.0]) colors in the three successive bins of increasing 24$\mu$m flux density are (0.70,0.64), (0.79,1.07) and (0.95,1.40) mag.  A two-sided Kolmogorov-Smirnov (hereafter K-S) test rules out the hypothesis that the color distribution in one of the three flux density bins is drawn from any other one at a very high level of significance.

\subsubsection{Overall Spectral Energy Distributions}

\begin{figure}[t]
\epsscale{0.73}
\plotone{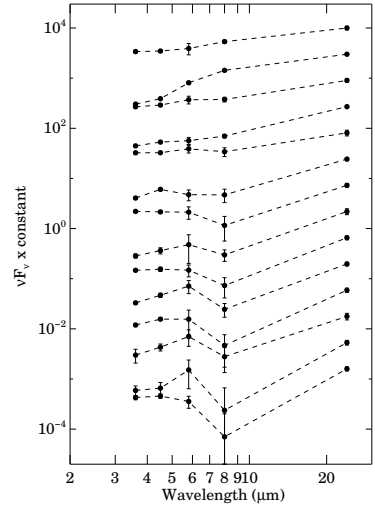}
\caption{Examples of the mid-infrared spectral energy distributions observed for the $(R-[24])\ge 14$ sample. The SEDs range from being power-law dominated (near the top) to exhibiting a distinct ``bump'' in their SED between 3--10$\mu$m (near the bottom). The ``bump'' is attributed to starlight in the galaxies, peaking at a rest-frame wavelength of $\approx$1.6$\mu$m.
\label{examples}}
\end{figure}

The SEDs of galaxies can provide important insights into their nature, i.e., the components that dominate the energetics at different wavelengths.  The overall SEDs of the sources with $(R-[24])\ge14$ span a wide range. At one extreme, the SEDs are fairly monotonic throughout the mid-IR bands, and are fairly well represented by a power-law, suggesting a broad range of (warm) dust temperatures similar to that observed in AGN-dominated sources. At the other extreme, the SEDs show a distinct ``bump'', which we interpret as being due to the presence of a stellar continuum peaking at a rest-frame wavelength of 1.6$\mu$m. Examples of the two classes of SEDs are shown in figure~\ref{examples}. There is a continuum of SEDs between these two extremes, and it is difficult to define an objective criterion that cleanly discriminates between these two classes, perhaps suggesting that the SEDs in a large fraction of the population are characterized by both AGN and star-formation activity.

In order to establish a more rigorous definition (more for statistical than physical purposes), we define the following objective means of classifying a source SED as being power-law dominated. We performed two power-law fits to the mid-infrared flux measurements of every source, the first to just the IRAC measurements (i.e., the 3.6$\mu$m to 8.0$\mu$m data) and the second to the combined IRAC and MIPS 24$\mu$m data. We then examined the power-law indices ($\alpha_{\rm IRAC}$ and $\alpha_{\rm IRAC+24}$; $F_\nu\propto \lambda^\alpha$) and $\chi^2$ values of the resulting fits. An SED was classified as a `power-law' source if the power-law index $\alpha_{\rm IRAC}>3$, or if $\alpha_{\rm IRAC}>1$ and $\chi^2_{\rm IRAC}\le 1.5$, or if  $\alpha_{\rm IRAC}>\alpha_{\rm IRAC+24}$ and $\vert\alpha_{\rm IRAC}-\alpha_{\rm IRAC+24}\vert > 2\sqrt{\sigma^2_{\rm IRAC}+\sigma^2_{\rm IRAC+24}}$, where $\sigma_{\rm IRAC},\sigma_{\rm IRAC+24}$ are the formal fit uncertainties on the $\alpha$'s. Although this selection appears complex, it does separate the sources which show monotonic SEDs from the ones which show ``bumps'' or are otherwise different (due to being too noisy, for example).

\begin{figure}[t]
\epsscale{1.0}
\plotone{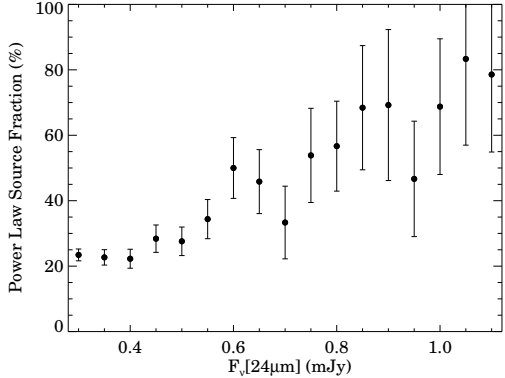}
\caption{The fraction of power-law sources as a function of 24$\mu$m flux density for the $(R-[24])\ge 14$ sample of galaxies. See text for details. 
\label{PLfrac}}
\end{figure}

Using this SED discriminant, we find that the spectral energy distributions are loosely correlated with 24$\mu$m flux density, with the bright $F_{\rm 24\mu m}$ sources being predominantly power-law sources (presumably with AGN emission dominating the mid-infrared spectral region) and the fainter $F_{\rm 24\mu m}$ sources exhibiting clearer `bumps' in their SEDs (Figure~\ref{PLfrac}). 

Our SED discriminant is constructed solely to capture this qualitative trend in the observed SEDs, and not to be rigorously used to  define unambiguous samples of `bump' or `power-law' sources. Different thresholds in $\chi^2$ or $\alpha$ in the criteria above change the fractions of power-law sources, but still reproduce the basic trend. 

The range of SEDs in our sample is similar to that observed by \citet[]{pol2006}. \citet[]{pol2006} define several classes based on the shape of the SED, and identify ``class I'' and ``class II'' sources as their primary sample of obscured infrared AGN. Our $(R-[24])\ge14$ criterion would select all of the ``class I'' sources and 36\% of the ``class II'' sources in the  \citet[]{pol2006} study. 

\subsection{Redshifts}

\begin{figure}[t]
\epsscale{0.9}
\plotone{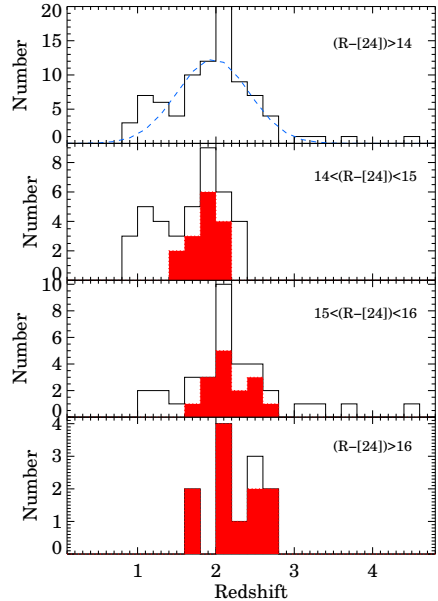}
\caption{{\it Top:} The observed redshift distribution of all \nz\ $(R-[24])\ge 14$ galaxies. The dashed line in the top panel shows the best fit Gaussian approximation to the observed redshift distribution, with $\bar{z}=1.99\pm 0.05, \sigma=0.45\pm 0.05$. {\it Lower three panels:} Redshift distributions in more restricted ranges of $(R-[24])$ color, as designated in the panel legends. The shaded histogram in each of the three lower panels shows the subset of 
redshifts measured from {\it Spitzer}/IRS observations.}
\label{zdist}
\end{figure}

The redshift histogram of all \nz\ galaxies is shown in Figure~\ref{zdist}. The distribution is broad, and can be modeled as a Gaussian centered at $\bar{z}=1.99\pm0.05$ with a dispersion $\sigma_z=0.45\pm0.05$. The overall redshift distribution is very similar to that of galaxy populations selected at sub-millimeter wavelengths \citep[the Sub-mm Galaxies, or SMGs;][]{cha2005}. 

It is noteworthy that the redshift distribution of galaxies with  $14\le (R-[24]) < 15$ differs statistically from that of redder galaxies with $(R-[24])\ge 15$. Both a Student's T-test and a Wilcoxon Rank-Sum test rule out the hypothesis that the two subsamples have similar means at $>$99.9\% significance, and a two-sided K-S test finds that the two distributions differ at the $\approx$99.6\% level of significance. The bulk of the redshifts for the reddest galaxies (i.e., $R-[24]\ge 15$) were obtained with {\it Spitzer/IRS}, but since the distributions of the IRS-determined redshifts also show a similar dependence on $R-[24]$ color, this observational bias does not appear to be the source of the difference (see the red shaded histograms in Figure~\ref{zdist}). Unfortunately the redshift distribution is too broad, and the spectroscopic redshifts too few, to trust any correlation between the $(R-[24])$ color and the redshift derived from the present data, but it does appear that our color selection criterion effectively selects against lower redshift sources. Indeed, applying the color criterion $(R-[24])\ge 14$(15) to the sample of \citet[]{yan2007} effectively selects only the sources with redshifts $z>1$(1.6).

\begin{figure}[t]
\epsscale{0.8}
\plotone{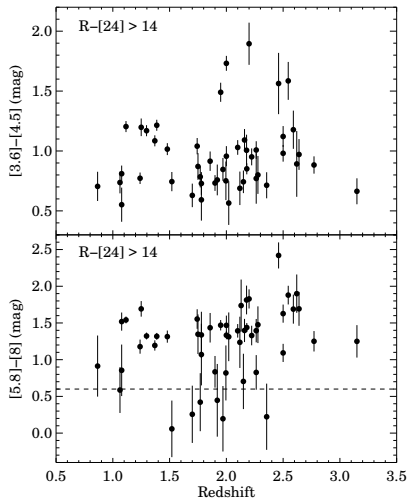}
\caption{The mid-infrared ${\rm [3.6\mu m]-[4.5\mu m]}$ (top panel) and ${\rm [5.8\mu m]-[8.0\mu m]}$ (bottom panel) colors as a function of redshift for DOGs with measured spectroscopic redshifts and $\ge 3 \sigma$ measurements of flux densities in the IRAC bands. The dashed line in the lower panel represents the lower color cut used by \citet[]{ste2005} to select AGN. There is no strong correlation between color and redshift for the $R-[24]\ge14$ sample. We interpret the large spread in color at a given redshift as representing a range in the relative contributions of reprocessed AGN and stellar luminosity to the observed mid-infrared bands, with $L_{\rm MIR}(AGN)/L_{\rm MIR}(stars)$ increasing toward redder (i.e., warmer) colors (cf.~Figure 11 of Mazzarella et al. 1992). Rather than a clean distinction between AGN and star-forming galaxies, there is a continuum of sources across the range of color.}
\label{colorvsz}
\end{figure}

There is no obvious dependence of redshift on mid-infrared color or 24$\mu$m flux density within the DOG population. In particular, we find no significant correlation between the [3.6]$-$[4.5] or [5.8]$-$[8.0] color with redshift within our sample (Figure~\ref{colorvsz}).  A two-sided K-S test finds no significant difference in the redshift distributions for galaxies brighter than or fainter than $F_{\rm 24\mu m}=1$~mJy. There is some evidence that the small subset of 22 galaxies with $F_{\rm 24\mu m} < 0.6$mJy does show a narrower redshift distribution centered at slightly lower redshifts ($\bar{z}(F_{\rm 24\mu m}<0.6{\rm mJy})\approx 1.86$).  This trend is also reinforced when considering the redshift distribution of all 24$\mu$m sources (irrespective of color). \citet[]{des2007} present the redshift distribution of a sample of galaxies selected with 24$\mu$m flux densities $F_{\rm 24\mu m}\ge0.3$mJy, and demonstrate that there is general trend of increasing mean redshift with $(R-[24])$ color. However, our current sample of \nz\ redshifts for $(R-[24])\ge14$ sources is a small subset, and without a larger spectroscopic sample, we are loathe to adopt anything other than the simplest assumptions regarding the redshift distribution. Since the current set of spectroscopic redshifts, although obtained in diverse ways, span the full range of optical magnitude and 24$\mu$m flux density of the overall DOG sample (see Figure~\ref{zspecDOGs}), we adopt the Gaussian fit to the observed redshift distribution as the overall redshift distribution of the sample. 

The redshift distribution implies that the observed $R-[24]$ selection corresponds (approximately) to a rest-frame 8$\mu$m to 2200\AA\ flux density ratio. In order to search for local counterparts of these galaxies, we examined the {\it Spitzer}/IRAC to {\it GALEX}/NUV flux density ratios of all 550 low-redshift ($z\le 0.1$) galaxies in Bo\"otes with spectroscopic redshifts from the AGES survey (Kochanek et al., in prep); none were found to have flux density ratios similar to those in our DOG sample (S. Salim, personal communication). 

\subsection{Rest-Frame Luminosities}

\subsubsection{8$\mu$m Luminosities}

The rest-frame 8$\mu$m luminosity, $L_8$, of the DOGs is estimated using 
\begin{eqnarray}
L_8\equiv\nu L_\nu\vert_{\rm 8\mu m} = \nu_{\rm 8\mu m}{{4\pi d_L^2}\over{1+z}} F_{\rm 8\mu m}\\
F_{\rm 8\mu m}\approx F_{\rm 24\mu m} \left({{1+z}\over{3}}\right)^\alpha,
\end{eqnarray}
where $\alpha\equiv 2.096~{\rm log}(F_{\rm 24\mu m}/F_{\rm 8\mu m})$ is the spectral index between 24$\mu$m and 8$\mu$m, $d_L$ is the luminosity distance, and $\nu_{\rm 8\mu m}$ is the frequency corresponding to 8$\mu$m. 

\begin{figure}[t]
\epsscale{1.0}
\plotone{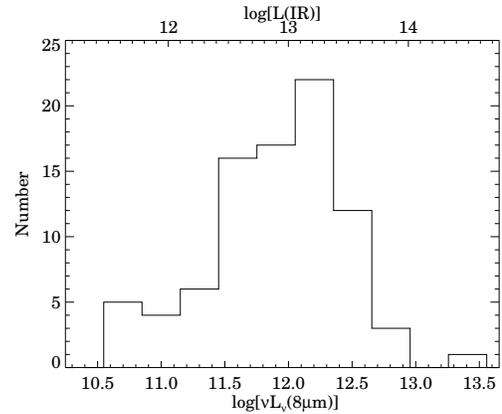}
\caption{Distribution of rest-frame 8$\mu$m luminosity ($\nu L_\nu({\rm 8\mu m})$) of all \nz\ sources in the $(R-[24])\ge14$ samples with spectroscopically measured redshifts. The top abscissa shows the corresponding $L_{\rm IR}$, estimated using the conversion of \citet[]{cap2007}. The bulk of the sources are clearly ULIRGs. (Our adopted conversion between $L_8$ and $L_{\rm IR}$ is only accurate to within a factor of three; to use a more [less] conservative conversion of $L_{\rm IR}=5 [15] L_8$, add $\approx 0.70$ [1.18] to the lower abscissa.) 
\label{Ldist}}
\end{figure}

The 8$\mu$m luminosities for the DOGs with measured spectroscopic redshifts are 
shown in figure~\ref{Ldist}.  Our estimates assume that the spectral shape between (observed frame) 8$\mu$m and 24$\mu$m is well matched by a power-law; this may be true if the mid-infrared emission from the galaxies is dominated by a featureless AGN, but not if the spectrum is strongly affected by PAH emission or silicate absorption. 

The derived $L_8$ values are large, with most of the galaxies being more luminous than $10^{11}\Lsun$ and half with $L_8\ge 10^{12}\Lsun$ ; by comparison, the 8$\mu$m luminosity function derived by \citet[]{red2007} for $z\approx 2$ UV-selected star-forming galaxies only extends to $\sim10^{11}\Lsun$. The median 24$\mu$m flux density of the sources with spectroscopic redshifts is ${\bar{F}}_{\rm 24\mu m}=1.04$mJy, in comparison with the entire sample, which has ${\bar{F}}_{\rm 24\mu m}\approx 0.4$. Hence, if the overall population of galaxies with $R-[24]\ge 14$ has the same redshift distribution as the subset of galaxies with spectroscopic redshifts, then the median rest-frame 8$\mu$m luminosity of the overall population is ${\bar{L}}_8\approx 4\times 10^{11}\Lsun$. 

\subsubsection{Far-Infrared Luminosities}

The distribution of far-infrared luminosities ($L_{\rm IR}$) for the DOGs with spectroscopically measured redshifts is shown in Figure~\ref{Ldist} (see upper abscissa). Nearly all (80 / \nz, or 93\%) of the sources  with spectroscopic redshifts are ULIRGs, with $L_{\rm IR} \ge 10^{12}\Lsun$, and more than half of these have $L_{\rm IR} \ge 10^{13}\Lsun$. 

Converting from observed 24$\mu$m flux densities to rest-frame far-infrared or bolometric luminosities is notoriously uncertain and has a checkered history \cite[e.g., see \S6 of][for a recent summary]{cap2007}. The conversion factors vary significantly depending upon the template adopted for the extrapolation, ranging from $L_{\rm IR}\approx 5~L_8$ for Mrk~231 to $L_{\rm IR}\approx 27.6~L_8$ for Arp~220. (Here, we adopt the \citet[]{san1996} definition of $L_{\rm IR}\equiv L({\rm 8\mu m - 1mm})$.) As pointed out by various authors \citep[e.g.,][]{red2007,cap2007}, many studies of the star-forming galaxy population have assumed the higher values, especially for the most 8$\mu$m-luminous galaxies. 
In constructing the upper abscissa of Figure~\ref{Ldist} we used the relation determined by \citet[]{cap2007} for the conversion, i.e., $L_{\rm IR}=1.91~L_8^{1.06}$, where $L_8\equiv \nu L_\nu\vert_{\rm 8\mu m}$ in units of $\Lsun$. However, this relation was derived by fitting [$L_8,L_{\rm IR}$] measurements of 93 low-redshift ($0< z \simlt 0.6$) galaxies (from the {\it Spitzer} First Look Survey), very few of which have $L_8 > 10^{11}\Lsun$ or $L_{\rm IR}>10^{12}$. Hence the extrapolation of this relation to higher redshift and higher luminosity galaxies  is questionable. 

Therefore, we also quote a range of $L_{\rm IR}$ values determined from $L_{\rm IR}= 10^{+5}_{-5}~L_8$. This range encompasses the conversion estimate based on Mrk~231 at the low end, and that adopted by \citet[]{red2006} for $z\sim 2$ galaxies at the high end. Since the mid-infrared emission from the galaxies in our sample appears to be dominated by AGN at the bright end and star-formation at the faint end, the large range we adopt should encompass the true values for $L_{\rm IR}$. Clearly, measurements of the rest-frame far-infrared spectral energy distributions of these galaxies are necessary before more precise estimates of $L_{\rm IR}$ can be made (cf.~Tyler et al. 2008, ApJ submitted).
The high far-infrared luminosities for DOGs hold even if we adopt the more conservative scaling based on Mrk~231: if $L_{\rm IR}=5~L_8$, then 75 / \nz, or 87\% of the sources with spectroscopic redshifts are ULIRGs.  


\subsubsection{Rest-Frame UV Luminosities}

\begin{figure}[t]
\epsscale{1.0}
\plotone{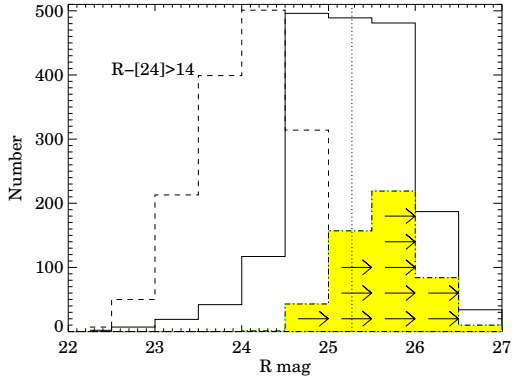}
\caption{Distribution of $R$-band (Vega) magnitudes for all the $(R-[24])\ge14$ galaxies (shown as the solid-line histogram) in the Bo\"otes Field, compared with the distribution for the $z\approx2$ ``BX'' UV-selected star-forming galaxies from \citet[]{red2007} (shown as the dashed-line histogram). The subset of our sample with only upper limits in the $R$-band are represented by the shaded histogram. The vertical dotted line represents the magnitude limit of the ``BX" sample. The $(R-[24])\ge14$ galaxy sample lies to fainter magnitudes than the ``BX" sample, with half the sample fainter than the magnitude limit of the ``BX" sample. The histograms shown are not normalized for the survey areas. }
\label{RbandHist}
\end{figure}

By selection, the galaxies in our sample are faint at optical wavelengths. Figure~\ref{RbandHist} shows the distribution of optical apparent magnitudes for the $(R-[24])\ge 14$ sample. Most of the galaxies are fainter than $R$=24.5 mag, and $\approx$27\% of the galaxies have only upper limits for the $R$-band magnitudes. For comparison, we also plot the distribution of $R$-band magnitudes for the $z\approx 2$ UV-selected star-forming galaxies from \citet[]{red2007} (selected using their ``BX" selection criteria). The majority of galaxies in our sample lie at fainter $R$-band magnitudes than the bulk of the ``BX" galaxy sample, with roughly half having magnitudes fainter than the magnitude limit of the ``BX" selection. 

\begin{figure}[t]
\epsscale{1.0}
\plotone{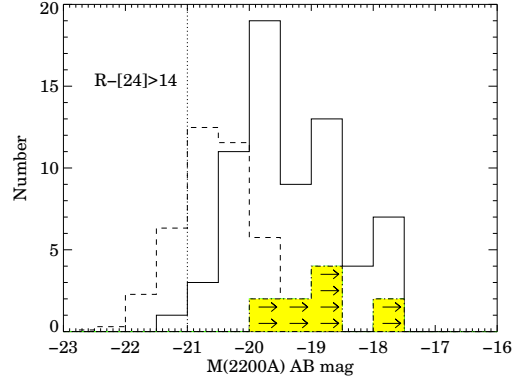}
\caption{The solid-line histogram shows the distribution of 2200\AA\ rest-frame absolute magnitudes (in AB units) of the $(R-[24])\ge14$ galaxies with spectroscopically measured redshifts. The subset of our sample with only upper limits in the $R$-band are represented by the shaded histogram. The absolute magnitudes of the $z\approx 2$ ``BX" UV-selected star-forming galaxies is shown by the dashed line histogram, and the vertical dotted line represents the magnitude of an $L^*$ galaxy. Nearly all the galaxies in the  $(R-[24])\ge14$ sample lie below $L^*$, and reach very faint rest-frame UV absolute magnitudes despite their remarkable mid-infrared luminosities. The histograms shown are not normalized for the survey areas.}
\label{2200Hist}
\end{figure}

Figure~\ref{2200Hist} shows the distribution of absolute magnitudes (at 2200\AA) for the subset of our sample with measured spectroscopic redshifts. Here, the absolute magnitude is estimated using $M_{\rm 2200\AA}=R_{\rm AB}-5{\rm log_{10}}(d_L/10{\rm parsec})-2.5{\rm log_{10}}(1+z)$. The median UV luminosity of our population is ${\bar{M}}_{\rm 2200\AA}\approx -19.7$. The distribution of UV luminosities for the ``BX'' sample of \citet[]{red2007} is also shown for comparison. Nearly all the $(R-[24])\ge14$ galaxies are fainter than $L^*$ \citep[$M^*\approx-21.0$AB mag, as determined from the ``BX'' sample of][]{red2007}, and many are approaching the UV rest-frame luminosities of local dwarf galaxies. Traditionally, low UV luminosity galaxies are considered to be less bolometrically luminous as well, and thought to contribute less to the overall IR luminosity budget. In contrast, the galaxies in our sample have low rest-frame UV luminosities but are nevertheless very bolometrically luminous and belong at the top end of the galaxy luminosity function. This underscores the importance of a multiwavelength approach in selecting and studying galaxy populations at high redshift. 

We have ignored the $k$-corrections in our conversion to absolute magnitude above, and assumed that all our sample galaxies lie at $z=2$. This is defensible for the flat-spectrum galaxies in the ``BX'' sample of \citet[]{red2007}, but perhaps less so for our galaxy sample, whose members have very red spectral energy distributions. Nevertheless, we have adopted this conversion for simplicity, and this choice makes little difference to the overall comparison of the two samples. 

\subsection{X-Ray Properties}


Observations at X-ray wavelengths are well suited to identifying AGN over a range of redshift.  The Bo\"otes Field of the NDWFS has been mapped in its entirety by the {\it Chandra X-ray Observatory} \citep[i.e., the XBo\"otes Survey;][]{ken2005,mur2005,bra2006a}.  In order to determine whether the population of galaxies with extreme $R-[24]$ colors is dominated by AGN or star-forming galaxies, we examined their X-ray emission. 

\begin{figure}[t]
\epsscale{1.0}
\plotone{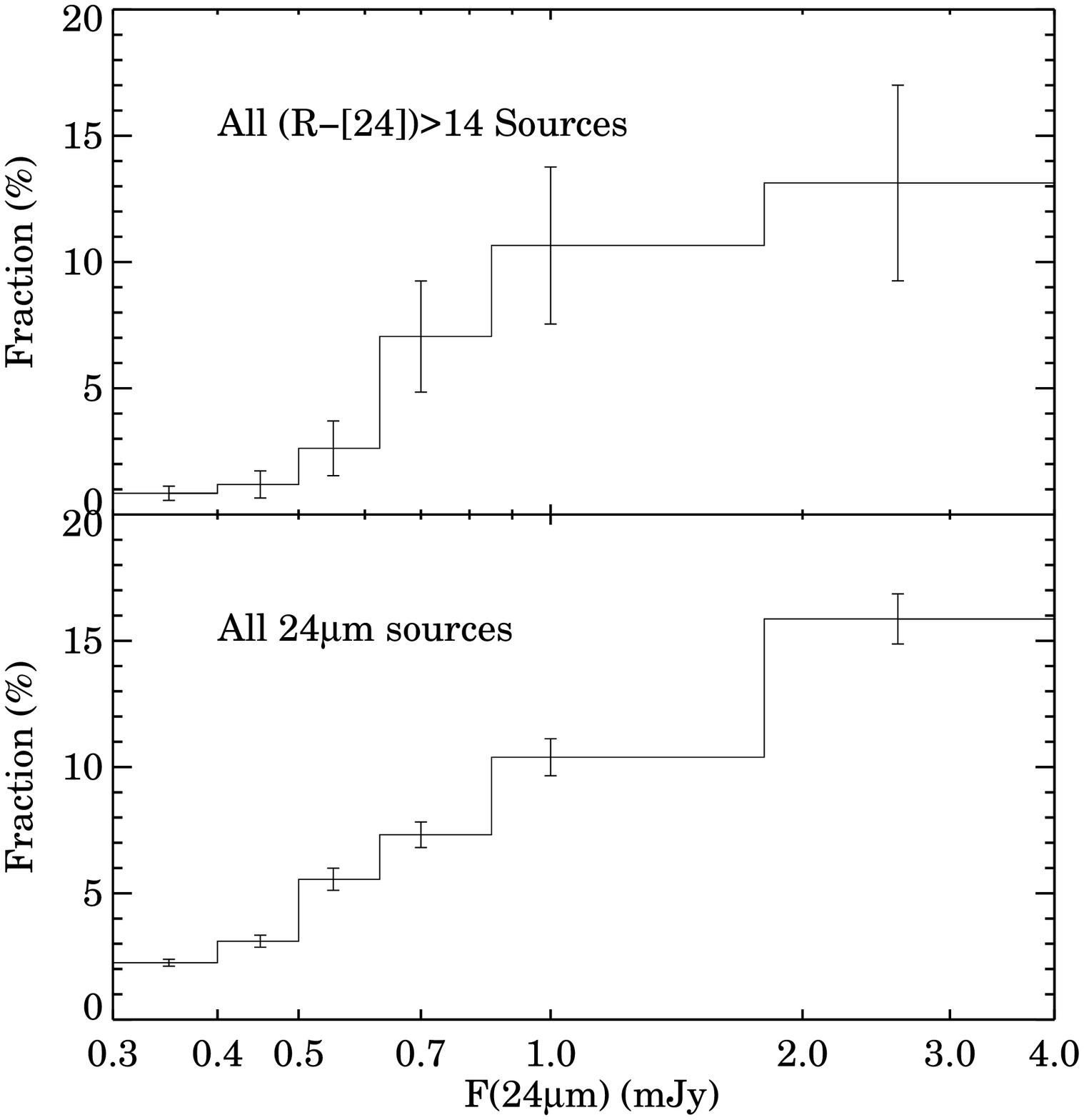}
\caption{{\it Top panel:} The fraction of galaxies in the $(R-[24])\ge 14$ sample detected as X-ray sources in the {\it Chandra} XBo\"otes survey \citep[]{mur2005,ken2005,bra2006a}, shown as a function of 24$\mu$m flux density.  Sources which are brighter at 24$\mu$m have a higher X-ray identification rate. {\it Bottom panel:} Fraction of all 24$\mu$m sources detected as X-ray sources in the XBo\"otes Survey. The X-ray identification fraction in the $(R-[24])\ge 14$ sample is similar to that of all 24$\mu$m sources at the bright end, but smaller at faint flux densities $F_{\rm 24\mu m}\sim 0.3-0.6$~mJy.
\label{Xrays}}
\end{figure}

At a redshift $z=2$, the {\it Chandra} soft (0.5$-$2~keV) and hard (2$-$7 keV) bands correspond to rest-frame energies of 1.5-6~keV and 6-21 keV respectively. 2,101 of the galaxies in our sample lie within $10'$ of the center of a {\it Chandra X-ray Observatory} ACIS observation in the XBo\"otes Survey. Of these galaxies, 59 have X-ray detections with $\ge$4 counts in the XBo\"otes survey; this corresponds to an X-ray flux of $F_{\rm 0.5-7keV}\approx 7.8\times10^{-15}\,{\rm erg\ cm^{-2}\ s^{-1}}$ and, at $z=2$, a luminosity of $L_X=2\times10^{44}\,{\rm erg\ s^{-1}}$ \citep[]{bra2006a}. This luminosity is more typical of AGN than of star-forming systems \citep[e.g.,][]{col2004}. As shown in the upper panel of figure~\ref{Xrays}, the X-ray detection rate of the galaxies in our sample increases with 24$\mu$m flux density, from $\approx 1\%$ at $F_{\rm 24\mu m}\approx 0.3$~mJy to $\simgt 10\%$ at $F_{\rm 24\mu m}\simgt 1$~mJy. This detection rate is similar to that of the overall 24$\mu$m source population (shown in the lower panel of Figure~\ref{Xrays}) at bright $F_{\rm 24\mu m}$, but a factor of two lower at the faint end, suggesting that the AGN fraction within the $(R-[24])\ge 14$ sample is an increasing function of 24$\mu$m flux density.

The X-ray data suggest that even the fainter galaxies may contain a significant AGN component, despite displaying oft-used discriminants of star-forming galaxies (e.g., stellar ``bumps'' in their SEDs, colors typically attributed to star-forming galaxies, low values of the 24$\mu$m to 8$\mu$m flux ratio, and mid-infrared spectra exhibiting PAH emission).  Although there may be AGN in the fainter sources, it is unclear from the current data whether the AGN plays a significant role in the mid/far-infrared luminosity of the fainter $F_{\rm 24\mu m}$ galaxies, but if it does, then the global star formation rate density derived by including the full contributions of their mid/far-infrared luminosity may be overestimates.

\section{Discussion}

\subsection{The Nature of the Extreme Red Population}

What are the extreme red objects that become an increasing fraction of the population at fainter 24$\mu$m flux densities? The spectroscopically measured redshifts for this population suggest prodigious luminosites ($10^{12-14}\Lsun$) and redder colors than any local ULIRGs.   We therefore suggest that the rise in this red fraction is predominantly an evolutionary effect due to the appearance of a more enshrouded population of luminous objects at higher redshift. The space density of this population is low; if the comoving space density remains constant with time, they may evolve into the fairly luminous local ellipticals. We speculate that we have uncovered a population of young galaxies undergoing very rapid bulge formation and / or AGN accretion, that evolve into fairly luminous galaxies ($\sim 4L^*$) in the present-day universe.

The optical-mid-IR SEDs of this population show a range of color. However, by selection, these objects are extreme in their optical-to-mid-IR properties and, as demonstrated by Figure~\ref{color}b, have colors redder than any local dusty galaxy or ULIRG. Since we have only limited spectral information, the dust reddening cannot be determined unambiguously. However, the existing mid-infrared spectroscopy shows that these galaxies have spectra that rise into the mid-infrared, suggesting emission from warm dust, and exhibit either deep silicate absorption \citep[e.g.,][]{hou2005,wee2006} or PAH emission \citep[][; Desai et al. 2008, in prep]{yan2007}. Near-infrared spectroscopy by our group for a handful of galaxies reveals, where measurable, large Balmer H$\alpha$/H$\beta$ decrements in both the narrow and broad components of the Balmer lines \citep[]{bra2007}. This suggests that the galaxies contain significant quantities of dust distributed on scales of, at the very least, the extended (i.e., kiloparsec scale) narrow-line emitting region. Since these galaxies lie at high redshift and are therefore very luminous, they must be undergoing significant AGN accretion and / or star-formation activity. This would suggest that the large rest-frame near-infrared to ultraviolet luminosity ratios result from heavy extinction of the most luminous emitting regions. 

\begin{figure}[t]
\epsscale{1.0}
\plotone{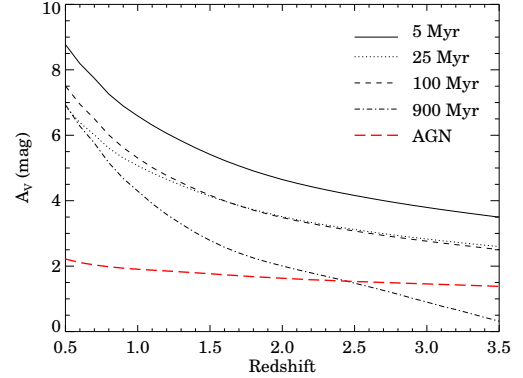}
\caption{The minimum extinction required to redden the spectral energy distributions of young stellar populations and AGN to produce an optical-to-mid-infrared color of $(R-[24])=14$. The template SEDs are models from the \citet[]{bc2003} population synthesis library with ages of 5~Myr (solid line), 25~Myr (dotted line), 100~Myr (short-dashed line) and 900~Myr (dot-dashed line), and the median QSO template (long-dashed red line) from \citet[]{elv1994}. The reddening law used is a combination of that from \citet[]{cal2000} and longer wavelength estimates from \citet[]{dra2003}, and assumes the unrealistic case of a dust screen in front of the emitting source in order to derive a firm lower limit on $A_V$.
\label{AV}}
\end{figure}

It is difficult to estimate the extinction in these galaxies from the limited photometric information, but we can derive some general constraints by considering the extinction necessary to create such red SEDs given a young stellar population. Figure~\ref{AV} shows the $V$-band extinction (in magnitudes) necessary to create an $(R-[24])=14$ color for template galaxies with ages less than 1~Gyr. This is based on the assumption of an extinction law derived using the functional form of \citet[]{cal2000} for $\lambda \le 2.2\mu$m and values of the extinction as tabulated in Table~4 of \citet[]{dra2003} at longer wavelengths. Although the assumption of a monolithic dust screen is unrealistic, it does provide a lower limit on the extinction required in these galaxies. In addition, the fact that the entire galaxy is so underluminous  at rest-frame UV wavelengths suggests that either the stellar population in these systems is old (i.e., $\simgt 1$~Gyr) and intrinsically faint at UV wavelengths, or that the dust is distributed on large enough scales to shroud the stellar component as well. If these galaxies are young and dominated by a star-forming phase, then they are enveloped on kiloparsec scales by prodigious amounts of dust. 

Given the redshifts ($z\sim2$) and spectral energy distributions of these galaxies, it is very unlikely that they are dominated by old stellar populations, and we therefore hypothesize that these objects are dust-obscured galaxies (DOGs), where most of the luminous regions of the galaxy are shrouded in dust. Such a scenario would suggest an overall SED for these galaxies containing three (perhaps related) components: a reddened stellar component peaking at a rest-frame wavelength of 1.6$\mu$m, cold dust enshrouding the star-forming regions reradiating the absorbed UV emission at long wavelengths (corresponding to, say, a rest-frame wavelength of 60$\mu$m), and warm dust enshrouding the AGN dominating the emission in the rest-frame mid-infrared (i.e., at rest-frame wavelengths of 5-20$\mu$m). 

If the luminosity from the sources selected by $(R-[24])\ge14$ is largely the result of star formation activity, then the implied star-formation rates are very large \citep[$\approx 1700 \Msun$/yr for an $L_{\rm IR}=10^{13}\Lsun$ object;][]{ken1998b}. Such a violent starburst is unlikely to be confined to a small region, and the (observed) very large infrared-to-ultraviolet luminosity ratio must result from dust distributed on very large (kiloparsec) scales.  In addition, star-formation rates this large cannot be sustained for long periods, suggesting that these galaxies are viewed during a short-lived phase during which the bulk of the stars is produced. 

If, instead, the sources are dominated by AGN, then the dust could be mostly confined to the nuclear regions of the galaxy. However, based on the measurement of large Balmer decrements in both broad- and narrow-line components in a few galaxies, \citet[]{bra2007} argue that the dust distribution must be patchy and over large volumes. Even if the galaxy luminosity is dominated by AGN activity, this phase of evolution must be fairly short lived. The bolometric luminosity of an AGN may be written (in terms of its Eddington Luminosity) as $L_{\rm bol}^{AGN}\equiv \epsilon L_{\rm Edd}\approx 0.33\times 10^{13} (\epsilon/0.1)(M_{\rm BH}/10^9\Msun) \Lsun$, or in terms of its accretion rate as $L_{\rm bol}^{AGN}=\eta \Mdot c^2 \approx 0.15\times 10^{13}(\eta/0.1)(\Mdot/1\Msun yr^{-1})\Lsun$. Large AGN luminosities therefore generally imply high accretion rates onto massive black holes. The present data on the DOGs do not allow any useful constraints on either the radiative efficiency, the accretion efficiency, or the black hole mass, but their luminosities of $10^{13}\Lsun$ suggest that they must be radiating at close to the Eddington luminosity even if they harbor a very massive black hole. Even at high accretion efficiencies (i.e., $\eta = 0.1$), the timescales to build up a $10^9\Msun$ black hole would be short ($\approx$0.15~Gyr). A flux density limited sample of galaxies (such as ours) will preferentially identify the galaxies in their most luminous phase, suggesting that we may be witnessing the last throes of the AGN formation in these obscured galaxies. 

What is the relevance of this rather extreme population to the overall galaxy population at $z\approx 2$? In the following subsections, we estimate the space density of the DOGs and their contribution to the infrared luminosity density in an attempt to address this question. In contrast to previous work, we do not discriminate between AGN and star-forming galaxies within our sample, but estimate their joint contribution to the space densities and contribution to the infrared luminosity density. 

\subsection{Space Density of DOGs}


Based on the observed redshift distribution, we now estimate the space density of this population. We assume that the comoving volume sampled by the $(R-[24])\ge 14$ and $F_{\rm 24\mu m}\ge 0.3$ selection criteria is simply the fraction of the volume between $0.5\le z \le 3.5$ weighted by the redshift distribution. In other words, the effective co-moving volume sampled is given by:
\begin{equation}
V_c^{\rm eff}({\rm DOGs}) = \Delta\Omega \int{g(z) dV_c}
\end{equation}
where $\Delta\Omega$ is the solid angle covered by the survey, 
\begin{equation}
g(z) = {{1}\over{\sigma_z\sqrt{2\pi}}} {\rm exp}\left[-{{(z-\bar{z})}\over{2\sigma_z^2}}\right]
\end{equation}
(with $\bar{z}=1.98$ and $\sigma_z=0.53$)
is the (normalized) Gaussian fit to the redshift distribution, and $dV_c$ is the comoving volume element per unit solid angle per unit redshift interval \citep[e.g.,][]{pee1993,hog1999}. For the Bo\"otes field, the volume sampled by the redshift distribution is roughly a third of the comoving volume between $0.5\le z \le 3.5$. The resulting average space density for this population in the redshift range $0.5\le z \le 3.5$ is 
\begin{equation}
\Sigma_{\rm DOG} (F_{\rm 24\mu m}\ge 0.3~{\rm mJy}) = (2.82\pm 0.05)\times 10^{-5} h_{70}^3~ {\rm Mpc^{-3}}. 
\end{equation}


This space density of DOGs is comparable to that of highly UV-luminous $z\sim 2-3$ star-forming galaxies selected using the Lyman-break or near-infrared selection techniques \citep[e.g.,][and references therein]{red2007,vand2006}. For example, the dust-obscured galaxy population has a space density comparable to that of $M_{\rm AB}({\rm 1700\AA})-5{\rm log}h_{0.7}\approx -22.2$ UV-bright galaxies at $z\sim 1.9-3.4$, based on the luminosity functions derived by \citet[]{red2007}. Although the space densities are comparable, it is important to note that each DOG contributes significantly more IR (and bolometric) luminosity per galaxy to the universe at $z\approx 2$. The population of ``massive galaxies'' selected as ``Distant Red Galaxies'' (DRGs) at near-infrared wavelengths has a space density of $(2.2\pm0.6)\times 10^{-4} h_{70}^3~{\rm galaxies~ Mpc^{-3}}$ \citep[]{vand2006}, roughly ten times more numerous than DOGs.  As mentioned above, the DRG and DOG populations also overlap in their rest-frame UV luminosity distributions, and it is therefore possible that a fraction of the star-forming DRGs are selected by our $(R-[24])\ge 14$ criterion. 

A recent measurement of the luminosity function by \citet[]{wak2006} finds that luminous (i.e., $>4L^*$) red galaxies in the local Universe have a space density of $\approx 2.5\times 10^{-5} h_{70}^3 {\rm Mpc^{-3}}$, very comparable to that of the $z\approx 2$ DOGs. It is tempting to suggest that DOGs may therefore represent a common phase of evolution for most massive galaxies. Indeed, if this phase is short lived (i.e., if the duty cycle of the enshrouded, luminous phase when galaxies may be selected as DOGs is short), then the DOGs may be the progenitors of a large fraction of present-day $>L^*$ galaxies. 

More detailed comparisons will require better determinations of the redshift distribution as a function of apparent magnitude and 24$\mu$m flux density. 

\subsection{Infrared Luminosity Density of DOGs}

In order to estimate the infrared luminosity density contributed by the DOGs, we begin by assuming, as before, that all the DOGs with no measured spectroscopic redshifts have a redshift probability distrbution described by the Gaussian approximation to the observed redshift distribution. For the DOGs with spectroscopic redshifts, we computed their $L_8$ values and converted them to $L_{\rm IR}$ estimates based on the procedure described above. For the DOGs without measured redshifts, we assigned random redshifts (drawn from the probability distribution) to each object and computed their luminosities. As in section \S3.3.2, we computed luminosities using the conversion of \citet[]{cap2007}, and derive a range based on the two extremes $L_{\rm IR}=[5-15]\times L_8$.  Since DOGs with fainter $F_{\rm 24\mu m}$ are also typically fainter in the IRAC bands, we adopt the mean $\bar{\alpha}=2.296$ as the spectral index for all sources. 

The infrared luminosity density (IRLD) of the entire sample was estimated using 1000 Monte Carlo simulations of the redshift distribution of the galaxies. We find the total infrared luminosity density contributed by the DOG population to be 
\begin{equation}
{\rm log[IRLD_{\rm DOG}(F_{\rm 24\mu m}\ge 0.3mJy)]} = 8.228^{+0.176}_{-0.302},
\end{equation}
where IRLD is in units of $\Lsun~{\rm Mpc^{-3}}$ and the limits are not 1$\sigma$ uncertainties, but the likely range of the IRLD based on the uncertainties in the $L_8$ to $L_{\rm IR}$ conversion (i.e., the lower and upper limits corresponding to $L_{\rm IR}/L_8= 5$ and 15 respectively). 

DOGs thus provide a significant contribution to the IRLD at $z\approx 2$. 
According to \citet[]{cap2007}, the total IRLD from all 24$\mu$m-selected $z\sim2$ galaxies is ${\rm log(IRLD_{\rm Total})}\approx 8.819^{+0.073}_{-0.079}$, with the contribution from ULIRGs alone being $42^{+15}_{-22}\%$ of the total.  Although DOGs presumably represent only a fraction of the $z\sim 2$ population (i.e., the reddest subset), they appear to contribute $\approx26\pm 14\%$ of the total IRLD from the 24$\mu$m-selected $z\sim 2$ galaxy population, and $\approx 60^{+40}_{-15}\%$ of the ${\rm IRLD_{\rm ULIRG}}$.  The \citet[]{cap2007} $z\sim2$ sample is identified (in the GOODS-N and -S fields) using photometric redshift estimates and should, in principle, include DOGs. What is noteworthy, however, is that the DOGs (selected solely on the basis of extreme $R-[24]$ color) comprise such a significant component of the IRLD$_{\rm ULIRG}$, which suggests that the DOG selection criterion effectively selects out the bulk of ULIRGs at $z\sim 2$. 

Not all $z\sim2$ galaxies may be selected as 24$\mu$m sources. Indeed, one of the most successful methods of selecting high-redshift galaxy populations is based on the UV emission of young, star-forming galaxies \citep[e.g.,][and references therein]{red2007}. Recently, \citet[]{red2007} have estimated the contribution of UV-selected galaxies to the IRLD at $z\sim2$. The IRLD contributed by DOGs is 158\%, 60\%, 35\% and 56\% of the IRLD contributed by the $L_{\rm IR}=10^{9-10}$, $10^{10-11}$, $10^{11-12}$, and $>10^{12}\Lsun$ UV-bright populations respectively. 

Do the bright and faint DOGs contribute comparably? To investigate this, we divided the DOG samples at $F_{\rm 24\mu m}=0.6$~mJy, resulting in 2,149 galaxies fainter than the cut with a median flux density of 0.37~mJy, and 452 galaxies brighter than the cut with a median flux density of 0.85mJy. The redshift distributions on either side of this cut are slightly different ($\bar{z}\approx1.9$/2.1 for the fainter/brighter galaxies with spectroscopically measured redshifts), but there are only 22 redshifts below the cut, so the distribution is not well constrained. If we assume that the distributions are the same at all flux densities, we find the IRLDs for the bright and faint subsamples are ${\rm log[IRLD_{\rm DOG}(F_{\rm 24\mu m}\ge 0.6)]=7.78\pm0.02}$ and  ${\rm log[IRLD_{\rm DOG}(0.3\le F_{\rm 24\mu m}< 0.6)]=7.91\pm0.01}$ respectively. The faint and bright sources thus contribute comparable amounts to the IRLD, with the increasing number making up for the lower luminosity of the fainter source population.

In actuality, there is a large range in optical-to-infrared color, and we have chosen the $(R-[24])\ge 14$ criterion simply to isolate the higher redshift galaxies. Since the redshift distribution is broad, selecting objects at a bluer threshold would still result in high-redshift population, but with a larger low-redshift contamination. Moreover, we recall that the number density of DOGs rises toward fainter 24$\mu$m flux densities, suggesting that $F_{\rm 24}<0.3$mJy sources may contribute significantly to the IRLD as well. In addition, we have not corrected our IRLD estimates for the sample incompleteness at the faint 24$\mu$m limit of our data. For all these reasons, the estimates of space density and luminosity density reported here for the DOGs are likely to be lower limits, corresponding to the contribution of only the most extreme (in $F_{\rm 24\mu m}$ and $(R-[24])$ color) subset of the population. It is therefore remarkable that even this relatively rare population can contribute such a significant fraction of the overall infrared luminosity density. 


\subsubsection{Comparison to Other $z\approx 2$ Galaxy Populations}

The galaxies selected by virtue of their extreme mid-infrared to optical flux density ratios seem to be under-represented in most optically selected samples of high-redshift galaxies. Indeed, our 24$\mu$m flux density criterion for DOGs (i.e., $F_{\rm 24\mu m}\ge 0.3$~mJy, imposed by the existing MIPS observations of the Bo\"otes Field) selects a relatively rare population that has only a handful of members in most existing small-field, deep samples. In this subsection, we attempt to compare our sample with other known $z\approx 2$ galaxy populations, most of which have been selected using observations at optical or near-infrared (i.e., rest-frame UV or optical) wavelengths.

As we mentioned in \S3.3.3 (and shown in Figures~\ref{RbandHist} and ~\ref{2200Hist}), the DOG population is systematically fainter than the UV-bright populations selected at $z\approx 2$ by, say, the ``BX/BM'' selection criteria of \citet[]{red2007}. This is not surprising, since we are selecting galaxies to have a very large mid-infrared to UV flux ratio. What is interesting is that our selection results in a sample that lies at high redshift, with no bright, low-redshift members. Moreover, it is important to note that the DOG population, while being faint at optical wavelengths, is actually bolometrically very luminous, and thereby contributes significantly to (and perhaps {\it dominates}) the top end of the bolometric luminosity function (i.e., at  $L_{\rm IR}\simgt 10^{12}\Lsun$). 


A comparison with the $z\approx 2-3$ ``Distant Red Galaxy" (DRG) samples selected using near-infrared selection techniques \citep[typically, $J-K>2.3$; e.g.,][]{franx2003,vand2004} is more difficult, since the Bo\"otes NDWFS field lacks very deep near-infrared imaging data. However, the DRG samples appear to have median $R_{\rm AB}\approx 25.9$ with 25(75)\%-ile values of the distribution at $R_{\rm AB}\approx 25.1(26.7)$ \citep[]{vand2006}. The median rest-frame luminosity of DRGs is $M_{\rm 2200\AA} \approx -18.3$~AB mag for $K<21$ (Vega) samples selected from the Hubble Deep Field South and MUSYC Deep fields \citep[G. Rudnick, personal communication;][]{gaw2006}. These magnitudes, magnitude ranges and rest-frame luminosities are comparable to those measured for our $(R-[24])\ge14$ population, suggesting that these populations may overlap. It is now understood that the DRG selection results in a mixed sample of passively evolving systems, star-forming galaxies and AGN \citep[e.g.,][]{kriek2006,kriek2007}. Since our selection criteria require galaxies to be bright at 24$\mu$m, it is likely that our selection does not include the passively evolving galaxies. 


Recently, much work has been done using the $BzK$ selection technique pioneered by \citet[]{dad2004} to select star-forming and passive galaxy populations at high redshift (i.e., $z\approx 2$). We examined the $(R-24)$ colors of the $BzK$ selected star-forming galaxies in the GOODS-N field \citep[]{dad2007a,dad2007b}, and find that only 4 of the 187 $BzK$ galaxies in GOODS-N satisfy our selection criteria. The vast majority of the $BzK$ galaxies are less extreme in their $(R-[24])$ colors, and are fainter in their 24$\mu$m flux densities. This may be due to the lower AGN contribution (by selection) to the mid-infrared luminosities of the $BzK$ star-forming population.  

Given the surface density of bright DOGs (i.e., with $F_{\rm 24\mu m}\ge 0.3$mJy, we would have expected roughly 14 galaxies in the GOODS-N field. Indeed, there are 21 galaxies with these selection criteria in the GOODS-N field, 3 of which are brighter than $F_{\rm 24\mu m}=1.0$~mJy (E. Daddi and M. Dickinson, personal communication). 
Fifteen of these 21 galaxies are also selected by the $BzK$ techniques, but all but 4 are rejected from the $BzK$ `star-forming' galaxy sample because they are hard X-ray sources (E. Daddi, personal communication).  The $(R-[24])\ge 14$ and $BzK$ selections therefore appear to be comparable, with the star-forming systems at bright 24$\mu$m flux densities effectively selected by both techniques. The simple $(R-[24])$ selection results in a slightly larger sample with perhaps a slightly broader redshift distribution. It is worth noting that the $BzK$ technique can be fairly expensive in telescope time, requiring deep imaging in three bands. The simple technique described here of using the $R$-[24] color selects similar star-forming galaxy and AGN populations more economically (90~sec with {\it Spitzer}/MIPS and 6000~sec of $R$-band per field). 

Recently, a similar population of dust-obscured galaxies was identified by \citet[]{yan2007} using the following selection criteria: $F_{\rm 24\mu m}>0.9$~mJy; $(\nu F_\nu({\rm 24\mu m})/\nu F_\nu({\rm 8\mu m}) > 3.16; (\nu F_\nu({\rm 24\mu m})/\nu F_\nu({\rm 0.7\mu m}) > 10$. Our selection criterion for DOGs corresponds to \\ $(\nu F_\nu({\rm 24\mu m}) / \nu F_\nu({\rm 0.7\mu m}) > 28.6$, and thus selects the redder, and a higher redshift subset of the population. The \citet[]{yan2007} galaxies with $(R-[24])\ge 14$(15) all lie at $z > 1$(1.6). The $(\nu F_\nu({\rm 24\mu m})/\nu F_\nu({\rm 8\mu m})$ distribution for DOGs shows that the selection criteria used by \citet[]{yan2007} would reject 36\% of the DOGs. 

Finally, although dust-obscured galaxies have been found in other studies using comparable selection criteria, we emphasize the simplicity and robustness of the current approach. In particular, the simple selection criteria will make it easier, in principle, to quantify the selection function once a large sample of spectroscopic redshifts become available. One significant difference in the way we select the sample of DOGs is that we do not discriminate between AGN-dominated and star-formation-dominated systems. The mid-infrared colors of many of the DOGs suggest that they may indeed be dominated by AGN at bright 24$\mu$m flux densities (i.e., $F_{\rm 24\mu m}\simgt 0.8$). Although the majority of the fainter population does show stellar ``bumps'', PAH emission features in their mid-infrared spectra, and other evidence of starlight in their spectral energy distributions, their mid-infrared,  far-infrared, or bolometric luminosity may yet contain a significant contribution from an AGN. The combination of AGN and star-formation signatures in the DOG population (and in some cases in the very same objects) suggests that the population may be undergoing both rapid black-hole growth and rapid star-formation. 

\subsection{The Possible Role of DOGs in Galaxy Evolution}

We have demonstrated that the simple selection criterion of large mid-infrared to optical flux density ratio efficiently isolates a significant population of $z\approx 2$ galaxies. These systems are extremely luminous at rest-frame mid-infrared wavelengths, presumably due to warm dust heated by some combination of AGN and hot, young stars. What is the role of these galaxies in galaxy evolution?

In a seminal paper, \citet[]{san1988} proposed that local ULIRGs represented a brief dust-enshrouded stage in the formation of quasars. Despite the extreme colors of the DOGs and the lack of local galaxies (even ULIRGs) with similar colors, there are some similarities between the two populations. They are both at the extreme luminous end of the galaxy population at their epochs, and they both appear to be heavily enshrouded systems exhibiting evidence for both intense star formation and AGN accretion activity. 

We therefore speculate that the DOGs represent a fairly short-lived phase in the evolution of massive (i.e., $\sim4L^*$) galaxies, in between the phases represented by the sub-millimeter galaxy population (SMGs) and the more ``passive'' galaxies (e.g., selected as passive $BzK$s or $DRG$s). We envision a scenario in which gas accumulates in deep potentials, triggering an early episode of star-formation which quickly results in the formation of large quantities of dust. This results in a system which is luminous at sub-mm wavelengths (as an SMG), with cold dust temperatures. The mid-infrared spectra of these systems would show PAH features, characteristic of star-forming systems, as has been observed \citep[e.g.,][]{men2007,pop2008}. At some point, as the accretion and star-formation proceed, an AGN is triggered which heats the dust to warmer temperatures. It is at this point that the system would have a warmer characteristic dust temperature and be selected as a DOG. If the AGN is able to destroy or expel most of the dust (or the star-formation proceeds to the point where the dust and gas are sufficiently consumed), the DOG would evolve into an optically visible AGN. Whether or not the DOG is transformed into an optically-luminous AGN will depend on the relative timescales of dust destruction / consumption and AGN accretion. It has been suggested recently that AGN feedback and star-burst driven winds can provide two important mechanisms for terminating the star-formation in galaxies \citep[e.g.,][and references therein]{hop2007b}. In such a scenario, the galaxy evolutionary phase represented by DOGs would follow the bulge growth phase (i.e., the SMGs?), but precede the phase when the galaxy may be visible as an optical quasar or, eventually, as a red, passively-evolving galaxy (i.e., the DRGs or passive $BzK$ galaxies?). If, instead, the AGN phase is more short lived than the UV-bright star-forming phase, then DOGs would evolve into a UV-bright star-forming population. 

There is some circumstantial support for such a scenario from the measured space density and redshift distribution of the DOGs. The DOG surface density on the sky (of 0.089~arcmin$^{-2}$) is similar to that of the luminous sub-mm galaxy population with 850$\mu$m flux densities $F_{\rm 850\mu m} > 6$mJy \citep[]{cop2006}, and they have a comparable redshift distribution \citep[]{cha2005}. The 850$\mu$m sub-mm flux densities of the DOG population have not yet been measured, but the upcoming SCUBA-II Legacy survey should provide useful constraints. In the scenario described above, the DOGs are expected to have warmer dust temperatures than the SMGs, and it is indeed interesting that the detection rate of SMGs at MIPS wavelengths (24$\mu$m and 70$\mu$m) is low, with only $\simlt 40$\% of the SMGs in the GOODS-N field having 24$\mu$m flux densities $F_{\rm 24\mu m}\ge 0.3$mJy \citep[]{pop2008}. The space density of the brightest DOGs (with $F_{\rm 24\mu m}\ge 1$~mJy) is comparable to that of unobscured QSOs at comparable redshifts \citep[e.g.,][]{bro2006}. If these brighter sources are powered primarily by AGN, they represent a significant component of the accretion history of supermassive black holes, missed by traditional QSO surveys.  

Clustering analyses of the population may provide a clue to the possible evolutionary products of DOGs and the lifetime of the DOG phase relative to the SMG phase, and we will investigate this in a future paper.  

\section{Summary}

KPNO $R$-band and {\it Spitzer} MIPS 24$\mu$m surveys of the Bo\"otes Field of the NOAO Deep Wide-Field Survey reveal a population of galaxies with very extreme optical to mid-infrared colors. Galaxies with such red colors are absent from the local universe; e.g., there are no galaxies with these extreme (rest-frame) flux density ratios among the $z<0.1$ galaxies in the Bo\"otes Field. From the 8.14~deg$^2$ 24$\mu$m/$R$-band overlap area, we select a sample of 2,603 galaxies with 24~$\mu$m flux densities $F_{\rm 24\mu m}\ge 0.3$~mJy and $(R-[24])\ge 14$ Vega mag (corresponding to $F_{\rm 24\mu m}/F_{0.7\mu m}\simgt 1000$. These galaxies increase in number and as a fraction of the total number towards fainter 24$\mu$m flux densities, rising from 7$\pm$0.6\% of the population at $F_{24\rm \mu m} \ge 1$mJy to 13$\pm$1\% at $F_{24\rm \mu m} \approx 0.3$mJy. This reddening of the overall source population and the rise of the red fraction is not due to a simple $k$-correction; the colors of this population of objects are redder than any local ULIRGs, suggesting that these are very dust enshrouded objects which constitute an increasing fraction of the source population to fainter 24$\mu$m flux densities. 

Selection of the reddest sources using a simple color cut of $(R-[24])\ge 14$ effectively isolates a high redshift population. The spectral energy distributions of these objects range from power-law SEDs suggestive of highly obscured AGN to SEDs exhibiting a ``bump'' at mid-infrared wavelengths. The fraction of sources with ``bump'' SEDs increases toward fainter 24$\mu$m flux densities. We interpret the ``bump'' in the SEDs of the latter group as arising from a stellar component peaking at a rest-frame wavelength of 1.6$\mu$m. 

Using spectrographs on the {\it Spitzer Space Telescope} and the telescopes of W. M. Keck Observatory, we have measured spectroscopic  redshifts for \nz\ galaxies and have determined that the majority of this population lies between the redshift range $1.5\ltsim z \ltsim 2.5$. The redshift distribution may be modeled as a Gaussian with a mean redshift of 1.99$\pm$0.05 and width $\sigma=0.45\pm0.05$. The high redshifts imply prodigious luminosities for this population. Corrections to bolometric luminosities are uncertain, but crude estimates suggest luminosites of $\sim 10^{12-14}\Lsun$. Although the dominant source of the luminosity is not conclusively established, it is very likely that both AGN and prodigious star-formation contribute to the luminosity. These dust-obscured galaxies show a range in mid-infrared color, probably due to a change in the relative dominance of AGN over star formation contributions to the mid-infrared luminosity. A rough estimate of the space density of this population suggests a comoving space density of $2.82\times 10^{-5}~h_{70}^3{\rm Mpc^{-3}}$. This dust obscured galaxy population contributes a significant fraction of the infrared luminosity density at $z\approx 2$, constituting roughly ${\rm log~IRLD \approx 8.228+0.176-0.302}$, where IRLD is in units of $\Lsun~{\rm Mpc^{-3}}$. This is roughly 26$\pm$14\% of the total IRLD contributed by galaxies at this redshift, and $60^{+40}_{-15}$\% of the IRLD contributed by luminous infrared galaxies with $L_{\rm IR}\ge10^{12}\Lsun$. 

The properties of this population suggest that they may be galaxies caught in the very process of formation, undergoing rapid growth of their stellar components and / or nuclear black holes. We speculate that these dust-obscured galaxies (or DOGs) may represent an evolutionary phase in the formation of massive galaxies, when the AGN is turned on and begins the process of terminating star-formation. Such a picture would envision DOGs as being the intermediate stage between the sub-mm galaxy population and the more quiescent galaxies (e.g., the `passive' DRGs). 

\acknowledgments

This work is based in part on observations made with the {\it Spitzer Space Telescope}, which is operated by the Jet Propulsion Laboratory, California Institute of Technology under NASA contract 1407. We are grateful to the expert assistance of the staff of Kitt Peak National Observatory where the Bo\"otes field observations of the NDWFS were obtained. The authors thank NOAO for supporting the NOAO Deep Wide-Field Survey. In particular, we thank Jenna Claver, Lindsey Davis, Alyson Ford, Emma Hogan, Tod Lauer, Lissa Miller, Erin Ryan, Glenn Tiede and Frank Valdes for their able assistance with the NDWFS data. AD thanks Naveen Reddy, Greg Rudnick, Mark Dickinson, and Samir Salim for illuminating discussions about the innumerable $z\approx2$ galaxy populations and local UV-bright populations, and Naveen Reddy, Alex Pope, Casey Papovich and the anonymous referee for constructive comments on the manuscript. AD is grateful to the Institute for Astronomy of the University of Hawaii, the Spitzer Science Center, the University of California at Santa Cruz, and Steward Observatory for their hospitality during his sabbatical year, and for enabling the writing of this paper. The research activities of AD and BTJ are supported by NOAO, which is operated by the Association of Universities for Research in Astronomy (AURA) under a cooperative agreement with the National Science Foundation.  Support for E. LeFloc'h was provided by NASA through the Spitzer Space Telescope Fellowship Program. We also thank the staffs of the W.~M.~Keck Observatory and the Gemini-North Observatory, where some of the galaxy redshifts were obtained. The Gemini Observatory is operated by the Association of Universities for Research in Astronomy, Inc., under a cooperative agreement with the NSF on behalf of the Gemini partnership: the National Science Foundation (United States), the Science and Technology Facilities Council (United Kingdom), the National Research Council (Canada), CONICYT (Chile), the Australian Research Council (Australia), CNPq (Brazil) and SECYT (Argentina). The authors also wish to recognize and acknowledge the very significant cultural role and reverence that the summit of Mauna Kea has always had within the indigenous Hawaiian community. 


\begin{thebibliography}{71}
\expandafter\ifx\csname natexlab\endcsname\relax\def\natexlab#1{#1}\fi

\bibitem[{{Bertin} \& {Arnouts}(1996)}]{ber1996}
{Bertin}, E., \& {Arnouts}, S. 1996, \aaps, 117, 393

\bibitem[{{Blain} {et~al.}(2002){Blain}, {Smail}, {Ivison}, {Kneib}, \&
  {Frayer}}]{bla2002}
{Blain}, A.~W., {Smail}, I., {Ivison}, R.~J., {Kneib}, J.-P., \& {Frayer},
  D.~T. 2002, \physrep, 369, 111

\bibitem[{{Brand} {et~al.}(2006){Brand}, {Brown}, {Dey}, {Jannuzi}, {Kochanek},
  {Kenter}, {Fabricant}, {Fazio}, {Forman}, {Green}, {Jones}, {McNamara},
  {Murray}, {Najita}, {Rieke}, {Shields}, \& {Vikhlinin}}]{bra2006a}
{Brand}, K., {et al.} 2006, \apj, 641, 140

\bibitem[{{Brand} {et~al.}(2007){Brand}, {Dey}, {Desai}, {Soifer}, {Bian},
  {Armus}, {Brown}, {Le Floc'h}, {Higdon}, {Houck}, {Jannuzi}, \&
  {Weedman}}]{bra2007}
{Brand}, K., {et al.} 2007, \apj, 663, 204

\bibitem[{{Brown} {et~al.}(2006){Brown}, {Brand}, {Dey}, {Jannuzi}, {Cool}, {Le
  Floc'h}, {Kochanek}, {Armus}, {Bian}, {Higdon}, {Higdon}, {Papovich},
  {Rieke}, {Rieke}, {Smith}, {Soifer}, \& {Weedman}}]{bro2006}
{Brown}, M.~J.~I., {et al.} 2006, \apj, 638, 88

\bibitem[{{Bruzual} \& {Charlot}(2003)}]{bc2003}
{Bruzual}, G., \& {Charlot}, S. 2003, \mnras, 344, 1000

\bibitem[{{Calzetti} {et~al.}(2000){Calzetti}, {Armus}, {Bohlin}, {Kinney},
  {Koornneef}, \& {Storchi-Bergmann}}]{cal2000}
{Calzetti}, D., {Armus}, L., {Bohlin}, R.~C., {Kinney}, A.~L., {Koornneef}, J.,
  \& {Storchi-Bergmann}, T. 2000, \apj, 533, 682

\bibitem[{{Caputi} {et~al.}(2007){Caputi}, {Lagache}, {Yan}, {Dole},
  {Bavouzet}, {Le Floc'h}, {Choi}, {Helou}, \& {Reddy}}]{cap2007}
{Caputi}, K.~I., {et al.} 2007, \apj, 660, 97

\bibitem[{{Chapman} {et~al.}(2005){Chapman}, {Blain}, {Smail}, \&
  {Ivison}}]{cha2005}
{Chapman}, S.~C., {Blain}, A.~W., {Smail}, I., \& {Ivison}, R.~J. 2005, \apj,
  622, 772

\bibitem[{{Chapman} {et~al.}(2004){Chapman}, {Smail}, {Windhorst}, {Muxlow}, \&
  {Ivison}}]{cha2004b}
{Chapman}, S.~C., {Smail}, I., {Windhorst}, R., {Muxlow}, T., \& {Ivison},
  R.~J. 2004, \apj, 611, 732

\bibitem[{{Chary} {et~al.}(2004){Chary}, {Casertano}, {Dickinson}, {Ferguson},
  {Eisenhardt}, {Elbaz}, {Grogin}, {Moustakas}, {Reach}, \& {Yan}}]{chary2004}
{Chary}, R., {et al.} 2004, \apjs, 154, 80

\bibitem[{{Chary} \& {Elbaz}(2001)}]{chary2001}
{Chary}, R., \& {Elbaz}, D. 2001, \apj, 556, 562

\bibitem[{{Colbert} {et~al.}(2004){Colbert}, {Heckman}, {Ptak}, {Strickland},
  \& {Weaver}}]{col2004}
{Colbert}, E.~J.~M., {Heckman}, T.~M., {Ptak}, A.~F., {Strickland}, D.~K., \&
  {Weaver}, K.~A. 2004, \apj, 602, 231

\bibitem[{{Coppin} {et~al.}(2006){Coppin}, {Chapin}, {Mortier}, {Scott},
  {Borys}, {Dunlop}, {Halpern}, {Hughes}, {Pope}, {Scott}, {Serjeant}, {Wagg},
  {Alexander}, {Almaini}, {Aretxaga}, {Babbedge}, {Best}, {Blain}, {Chapman},
  {Clements}, {Crawford}, {Dunne}, {Eales}, {Edge}, {Farrah}, {Gazta{\~n}aga},
  {Gear}, {Granato}, {Greve}, {Fox}, {Ivison}, {Jarvis}, {Jenness}, {Lacey},
  {Lepage}, {Mann}, {Marsden}, {Martinez-Sansigre}, {Oliver}, {Page},
  {Peacock}, {Pearson}, {Percival}, {Priddey}, {Rawlings}, {Rowan-Robinson},
  {Savage}, {Seigar}, {Sekiguchi}, {Silva}, {Simpson}, {Smail}, {Stevens},
  {Takagi}, {Vaccari}, {van Kampen}, \& {Willott}}]{cop2006}
{Coppin}, K., {et al.} 2006, \mnras, 372, 1621

\bibitem[{{Daddi} {et~al.}(2007{\natexlab{a}}){Daddi}, {Alexander},
  {Dickinson}, {Gilli}, {Renzini}, {Elbaz}, {Cimatti}, {Chary}, {Frayer},
  {Bauer}, {Brandt}, {Giavalisco}, {Grogin}, {Huynh}, {Kurk}, {Mignoli},
  {Morrison}, {Pope}, \& {Ravindranath}}]{dad2007b}
{Daddi}, E., {et al.} 2007{\natexlab{a}}, \apj, 670, 156

\bibitem[{{Daddi} {et~al.}(2004){Daddi}, {Cimatti}, {Renzini}, {Fontana},
  {Mignoli}, {Pozzetti}, {Tozzi}, \& {Zamorani}}]{dad2004}
{Daddi}, E., {et al.} 2004, \apj, 617, 746

\bibitem[{{Daddi} {et~al.}(2007{\natexlab{b}}){Daddi}, {Dickinson}, {Morrison},
  {Chary}, {Cimatti}, {Elbaz}, {Frayer}, {Renzini}, {Pope}, {Alexander},
  {Bauer}, {Giavalisco}, {Huynh}, {Kurk}, \& {Mignoli}}]{dad2007a}
{Daddi}, E., {et al.} 2007{\natexlab{b}}, \apj, 670, 173

\bibitem[{{Desai} {et~al.}(2007){Desai}, {Soifer}, {Dey}, {Jannuzi}, {Le
  Floc'h}, {Bian}, {Brand}, {Brown}, {Armus}, {Weedman}, {Houck}, {Cool}, \&
  {Stern}}]{des2007}
{Desai}, V., {et al.} 2007, \apj, 999, 999

\bibitem[{{Dey} {et~al.}(1999){Dey}, {Graham}, {Ivison}, {Smail}, {Wright}, \&
  {Liu}}]{dey1999}
{Dey}, A., {Graham}, J.~R., {Ivison}, R.~J., {Smail}, I., {Wright}, G.~S., \&
  {Liu}, M.~C. 1999, \apj, 519, 610

\bibitem[{{Draine}(2003)}]{dra2003}
{Draine}, B.~T. 2003, \araa, 41, 241

\bibitem[{{Eisenhardt} {et~al.}(2004){Eisenhardt}, {Stern}, {Brodwin}, {Fazio},
  {Rieke}, {Rieke}, {Werner}, {Wright}, {Allen}, {Arendt}, {Ashby}, {Barmby},
  {Forrest}, {Hora}, {Huang}, {Huchra}, {Pahre}, {Pipher}, {Reach}, {Smith},
  {Stauffer}, {Wang}, {Willner}, {Brown}, {Dey}, {Jannuzi}, \&
  {Tiede}}]{eis2004}
{Eisenhardt}, P.~R., {et al.} 2004, \apjs, 154, 48

\bibitem[{{Elbaz} {et~al.}(1999){Elbaz}, {Cesarsky}, {Fadda}, {Aussel},
  {D{\'e}sert}, {Franceschini}, {Flores}, {Harwit}, {Puget}, {Starck},
  {Clements}, {Danese}, {Koo}, \& {Mandolesi}}]{elb1999}
{Elbaz}, D., {et al.} 1999, \aap, 351, L37

\bibitem[{{Elvis} {et~al.}(1994){Elvis}, {Wilkes}, {McDowell}, {Green},
  {Bechtold}, {Willner}, {Oey}, {Polomski}, \& {Cutri}}]{elv1994}
{Elvis}, M., {et al.} 1994,
  \apjs, 95, 1

\bibitem[{{Faber} {et~al.}(2002)}]{deimos}
{Faber}, S., {et~al.} 2002, Proc. SPIE, 4841, 1657

\bibitem[{{Fabricant} {et~al.}(2005){Fabricant}, {Fata}, {Roll}, {Hertz},
  {Caldwell}, {Gauron}, {Geary}, {McLeod}, {Szentgyorgyi}, {Zajac}, {Kurtz},
  {Barberis}, {Bergner}, {Brown}, {Conroy}, {Eng}, {Geller}, {Goddard},
  {Honsa}, {Mueller}, {Mink}, {Ordway}, {Tokarz}, {Woods}, {Wyatt}, {Epps}, \&
  {Dell'Antonio}}]{hectospec}
{Fabricant}, D., {et al.} 2005, \pasp, 117, 1411

\bibitem[{{Fazio} {et~al.}(2004){Fazio}, {Hora}, {Allen}, {Ashby}, {Barmby},
  {Deutsch}, {Huang}, {Kleiner}, {Marengo}, {Megeath}, {Melnick}, {Pahre},
  {Patten}, {Polizotti}, {Smith}, {Taylor}, {Wang}, {Willner}, {Hoffmann},
  {Pipher}, {Forrest}, {McMurty}, {McCreight}, {McKelvey}, {McMurray}, {Koch},
  {Moseley}, {Arendt}, {Mentzell}, {Marx}, {Losch}, {Mayman}, {Eichhorn},
  {Krebs}, {Jhabvala}, {Gezari}, {Fixsen}, {Flores}, {Shakoorzadeh}, {Jungo},
  {Hakun}, {Workman}, {Karpati}, {Kichak}, {Whitley}, {Mann}, {Tollestrup},
  {Eisenhardt}, {Stern}, {Gorjian}, {Bhattacharya}, {Carey}, {Nelson},
  {Glaccum}, {Lacy}, {Lowrance}, {Laine}, {Reach}, {Stauffer}, {Surace},
  {Wilson}, {Wright}, {Hoffman}, {Domingo}, \& {Cohen}}]{faz2004}
{Fazio}, G.~G., {et al.}  2004, \apjs, 154, 10

\bibitem[{{Franceschini} {et~al.}(2001){Franceschini}, {Aussel}, {Cesarsky},
  {Elbaz}, \& {Fadda}}]{fra2001}
{Franceschini}, A., {Aussel}, H., {Cesarsky}, C.~J., {Elbaz}, D., \& {Fadda},
  D. 2001, \aap, 378, 1

\bibitem[{{Franx} {et~al.}(2003){Franx}, {Labb{\'e}}, {Rudnick}, {van Dokkum},
  {Daddi}, {F{\"o}rster Schreiber}, {Moorwood}, {Rix}, {R{\"o}ttgering}, {van
  de Wel}, {van der Werf}, \& {van Starkenburg}}]{franx2003}
{Franx}, M., {et al.} 2003, \apjl, 587, L79

\bibitem[{{Gawiser} {et~al.}(2006){Gawiser}, {van Dokkum}, {Herrera}, {Maza},
  {Castander}, {Infante}, {Lira}, {Quadri}, {Toner}, {Treister}, {Urry},
  {Altmann}, {Assef}, {Christlein}, {Coppi}, {Dur{\'a}n}, {Franx}, {Galaz},
  {Huerta}, {Liu}, {L{\'o}pez}, {M{\'e}ndez}, {Moore}, {Rubio}, {Ruiz}, {Toft},
  \& {Yi}}]{gaw2006}
{Gawiser}, E., {et al.} 2006, \apjs, 162, 1

\bibitem[{{Gruppioni} {et~al.}(2005){Gruppioni}, {Pozzi}, {Lari}, {Oliver}, \&
  {Rodighiero}}]{gru2005}
{Gruppioni}, C., {Pozzi}, F., {Lari}, C., {Oliver}, S., \& {Rodighiero}, G.
  2005, \apjl, 618, L9

\bibitem[{{Hogg}(1999)}]{hog1999}
{Hogg}, D.~W. 1999, ArXiv Astrophysics e-prints

\bibitem[{{Hopkins} {et~al.}(2007){Hopkins}, {Hernquist}, {Cox}, \&
  {Keres}}]{hop2007b}
{Hopkins}, P.~F., {Hernquist}, L., {Cox}, T.~J., \& {Keres}, D. 2007, ArXiv
  e-prints, 706

\bibitem[{{Houck} {et~al.}(2004){Houck}, {Roellig}, {van Cleve}, {Forrest},
  {Herter}, {Lawrence}, {Matthews}, {Reitsema}, {Soifer}, {Watson}, {Weedman},
  {Huisjen}, {Troeltzsch}, {Barry}, {Bernard-Salas}, {Blacken}, {Brandl},
  {Charmandaris}, {Devost}, {Gull}, {Hall}, {Henderson}, {Higdon}, {Pirger},
  {Schoenwald}, {Sloan}, {Uchida}, {Appleton}, {Armus}, {Burgdorf},
  {Fajardo-Acosta}, {Grillmair}, {Ingalls}, {Morris}, \& {Teplitz}}]{irs}
{Houck}, J.~R., {et al.} 2004, \apjs, 154, 18

\bibitem[{{Houck} {et~al.}(2005){Houck}, {Soifer}, {Weedman}, {Higdon},
  {Higdon}, {Herter}, {Brown}, {Dey}, {Jannuzi}, {Le Floc'h}, {Rieke}, {Armus},
  {Charmandaris}, {Brandl}, \& {Teplitz}}]{hou2005}
{Houck}, J.~R., {et al.} 2005, \apjl, 622, L105

\bibitem[{{Hu} \& {Ridgway}(1994)}]{hu1994}
{Hu}, E.~M., \& {Ridgway}, S.~E. 1994, \aj, 107, 1303

\bibitem[{{Jannuzi} \& {Dey}(1999)}]{ndwfs}
{Jannuzi}, B.~T., \& {Dey}, A. 1999, in ASP Conf. Ser. 191: Photometric
  Redshifts and the Detection of High Redshift Galaxies, 111

\bibitem[{{Kennicutt}(1998)}]{ken1998b}
{Kennicutt}, Jr., R.~C. 1998, \apj, 498, 541

\bibitem[{{Kenter} {et~al.}(2005){Kenter}, {Murray}, {Forman}, {Jones},
  {Green}, {Kochanek}, {Vikhlinin}, {Fabricant}, {Fazio}, {Brand}, {Brown},
  {Dey}, {Jannuzi}, {Najita}, {McNamara}, {Shields}, \& {Rieke}}]{ken2005}
{Kenter}, A., {et al.}, M. 2005, \apjs, 161, 9

\bibitem[{{Kriek} {et~al.}(2006{\natexlab{a}}){Kriek}, {van Dokkum}, {Franx},
  {F{\"o}rster Schreiber}, {Gawiser}, {Illingworth}, {Labb{\'e}}, {Marchesini},
  {Quadri}, {Rix}, {Rudnick}, {Toft}, {van der Werf}, \& {Wuyts}}]{kriek2006}
{Kriek}, M., {et al.} 2006{\natexlab{a}}, \apj, 645, 44

\bibitem[{{Kriek} {et~al.}(2006{\natexlab{b}}){Kriek}, {van Dokkum}, {Franx},
  {Illingworth}, {Coppi}, {Forster Schreiber}, {Gawiser}, {Labbe}, {Lira},
  {Marchesini}, {Quadri}, {Rudnick}, {Taylor}, {Urry}, \& {van der
  Werf}}]{kriek2007}
{Kriek}, M., {et al.} 2007, \apj, 669, 776

\bibitem[{{Lacy} {et~al.}(2004){Lacy}, {Storrie-Lombardi}, {Sajina},
  {Appleton}, {Armus}, {Chapman}, {Choi}, {Fadda}, {Fang}, {Frayer},
  {Heinrichsen}, {Helou}, {Im}, {Marleau}, {Masci}, {Shupe}, {Soifer},
  {Surace}, {Teplitz}, {Wilson}, \& {Yan}}]{lac2004}
{Lacy}, M., {et al.} 2004, \apjs, 154, 166

\bibitem[{{Lagache} {et~al.}(2003){Lagache}, {Dole}, \& {Puget}}]{lag2003}
{Lagache}, G., {Dole}, H., \& {Puget}, J.-L. 2003, \mnras, 338, 555

\bibitem[{{Lagache} {et~al.}(2004){Lagache}, {Dole}, {Puget},
  {P{\'e}rez-Gonz{\'a}lez}, {Le Floc'h}, {Rieke}, {Papovich}, {Egami},
  {Alonso-Herrero}, {Engelbracht}, {Gordon}, {Misselt}, \&
  {Morrison}}]{lag2004}
{Lagache}, G., {et al.} 2004, \apjs, 154, 112

\bibitem[{{Le Floc'h} {et~al.}(2005){Le Floc'h}, {Papovich}, {Dole}, {Bell},
  {Lagache}, {Rieke}, {Egami}, {P{\'e}rez-Gonz{\'a}lez}, {Alonso-Herrero},
  {Rieke}, {Blaylock}, {Engelbracht}, {Gordon}, {Hines}, {Misselt}, {Morrison},
  \& {Mould}}]{lef2005}
{Le Floc'h}, E., {et al.} 2005, \apj, 632, 169

\bibitem[{{Marleau} {et~al.}(2004){Marleau}, {Fadda}, {Storrie-Lombardi},
  {Helou}, {Makovoz}, {Frayer}, {Yan}, {Appleton}, {Armus}, {Chapman}, {Choi},
  {Fang}, {Heinrichsen}, {Im}, {Lacy}, {Shupe}, {Soifer}, {Squires}, {Surace},
  {Teplitz}, \& {Wilson}}]{marl2004}
{Marleau}, F.~R., {et al.} 2004, \apjs, 154, 66

\bibitem[{{McCarthy}(2004)}]{mcc2004a}
{McCarthy}, P.~J. 2004, \araa, 42, 477

\bibitem[{{McLean} {et~al.}(1998){McLean}, {Becklin}, {Bendiksen}, {Brims},
  {Canfield}, {Figer}, {Graham}, {Hare}, {Lacayanga}, {Larkin}, {Larson},
  {Levenson}, {Magnone}, {Teplitz}, \& {Wong}}]{nirspec}
{McLean}, I.~S., {et al.} 1998, in Presented at the Society of Photo-Optical Instrumentation
  Engineers (SPIE) Conference, Vol. 3354, Proc. SPIE Vol. 3354, p. 566-578,
  Infrared Astronomical Instrumentation, Albert M. Fowler; Ed., ed. A.~M.
  {Fowler}, 566--578

\bibitem[{{Men{\'e}ndez-Delmestre} {et~al.}(2007){Men{\'e}ndez-Delmestre},
  {Blain}, {Alexander}, {Smail}, {Armus}, {Chapman}, {Frayer}, {Ivison}, \&
  {Teplitz}}]{men2007}
{Men{\'e}ndez-Delmestre}, K., {et al.} 2007, \apjl, 655, L65

\bibitem[{{Monet} {et~al.}(1998){Monet}, {Canzian}, {Dahn}, {Guetter},
  {Harris}, {Henden}, {Levine}, {Luginbuhl}, {Monet}, {Rhodes}, {Riepe},
  {Sell}, {Stone}, {Vrba}, \& {Walker}}]{usnoA2}
{Monet}, D.~B.~A., {et al.} 1998,
  VizieR Online Data Catalog, 1252, 0

\bibitem[{{Murray} {et~al.}(2005){Murray}, {Kenter}, {Forman}, {Jones},
  {Green}, {Kochanek}, {Vikhlinin}, {Fabricant}, {Fazio}, {Brand}, {Brown},
  {Dey}, {Jannuzi}, {Najita}, {McNamara}, {Shields}, \& {Rieke}}]{mur2005}
{Murray}, S.~S., {et al.} 2005, \apjs, 161, 1

\bibitem[{{Oke} {et~al.}(1995){Oke}, {Cohen}, {Carr}, {Cromer}, {Dingizian},
  {Harris}, {Labrecque}, {Lucinio}, {Schaal}, {Epps}, \& {Miller}}]{lris}
{Oke}, J.~B., {et al.} 1995, \pasp, 107, 375

\bibitem[{{Papovich} {et~al.}(2004){Papovich}, {Dole}, {Egami}, {Le Floc'h},
  {P{\'e}rez-Gonz{\'a}lez}, {Alonso-Herrero}, {Bai}, {Beichman}, {Blaylock},
  {Engelbracht}, {Gordon}, {Hines}, {Misselt}, {Morrison}, {Mould},
  {Muzerolle}, {Neugebauer}, {Richards}, {Rieke}, {Rieke}, {Rigby}, {Su}, \&
  {Young}}]{pap2004}
{Papovich}, C., {et al.} 2004, \apjs, 154, 70

\bibitem[{{Peebles}(1993)}]{pee1993}
{Peebles}, P.~J.~E. 1993, {Principles of physical cosmology} (Princeton Series
  in Physics, Princeton, NJ: Princeton University Press, |c1993)

\bibitem[{{P{\'e}rez-Gonz{\'a}lez} {et~al.}(2005){P{\'e}rez-Gonz{\'a}lez},
  {Rieke}, {Egami}, {Alonso-Herrero}, {Dole}, {Papovich}, {Blaylock}, {Jones},
  {Rieke}, {Rigby}, {Barmby}, {Fazio}, {Huang}, \& {Martin}}]{per2005}
{P{\'e}rez-Gonz{\'a}lez}, P.~G., {et al.} 2005,
  \apj, 630, 82

\bibitem[{{Polletta} {et~al.}(2006){Polletta}, {Wilkes}, {Siana}, {Lonsdale},
  {Kilgard}, {Smith}, {Kim}, {Owen}, {Efstathiou}, {Jarrett}, {Stacey},
  {Franceschini}, {Rowan-Robinson}, {Babbedge}, {Berta}, {Fang}, {Farrah},
  {Gonz{\'a}lez-Solares}, {Morrison}, {Surace}, \& {Shupe}}]{pol2006}
{Polletta}, M.~d.~C., {et al.} 2006, \apj, 642, 673

\bibitem[{{Pope} {et~al.}(2008){Pope}, {Chary}, {Alexander}, {Armus},
  {Dickinson}, {Elbaz}, {Frayer}, {Scott}, \& {Teplitz}}]{pop2008}
{Pope}, A., {et al.} 2008, \apj, 675, XXX

\bibitem[{{Reddy} {et~al.}(2006){Reddy}, {Steidel}, {Fadda}, {Yan}, {Pettini},
  {Shapley}, {Erb}, \& {Adelberger}}]{red2006}
{Reddy}, N.~A., {et al.} 2006, \apj, 644, 792

\bibitem[{{Reddy} {et~al.}(2007){Reddy}, {Steidel}, {Pettini}, {Adelberger},
  {Shapley}, {Erb}, \& {Dickinson}}]{red2007}
{Reddy}, N.~A., {Steidel}, C.~C., {Pettini}, M., {Adelberger}, K.~L.,
  {Shapley}, A.~E., {Erb}, D.~K., \& {Dickinson}, M. 2007, ArXiv e-prints, 706

\bibitem[{{Rieke} {et~al.}(2004){Rieke}, {Young}, {Engelbracht}, {Kelly},
  {Low}, {Haller}, {Beeman}, {Gordon}, {Stansberry}, {Misselt}, {Cadien},
  {Morrison}, {Rivlis}, {Latter}, {Noriega-Crespo}, {Padgett}, {Stapelfeldt},
  {Hines}, {Egami}, {Muzerolle}, {Alonso-Herrero}, {Blaylock}, {Dole}, {Hinz},
  {Le Floc'h}, {Papovich}, {P{\' e}rez-Gonz{\' a}lez}, {Smith}, {Su},
  {Bennett}, {Frayer}, {Henderson}, {Lu}, {Masci}, {Pesenson}, {Rebull}, {Rho},
  {Keene}, {Stolovy}, {Wachter}, {Wheaton}, {Werner}, \& {Richards}}]{rie2004}
{Rieke}, G.~H., {et al.} 2004, \apjs, 154, 25

\bibitem[{{Sanders} \& {Mirabel}(1996)}]{san1996}
{Sanders}, D.~B., \& {Mirabel}, I.~F. 1996, \araa, 34, 749

\bibitem[{{Sanders} {et~al.}(1988){Sanders}, {Soifer}, {Elias}, {Madore},
  {Matthews}, {Neugebauer}, \& {Scoville}}]{san1988}
{Sanders}, D.~B., {Soifer}, B.~T., {Elias}, J.~H., {Madore}, B.~F., {Matthews},
  K., {Neugebauer}, G., \& {Scoville}, N.~Z. 1988, \apj, 325, 74

\bibitem[{{Smail} {et~al.}(2002){Smail}, {Ivison}, {Blain}, \&
  {Kneib}}]{sma2002}
{Smail}, I., {Ivison}, R.~J., {Blain}, A.~W., \& {Kneib}, J.-P. 2002, \mnras,
  331, 495

\bibitem[{{Soifer} {et~al.}(1987){Soifer}, {Neugebauer}, \& {Houck}}]{soi1987a}
{Soifer}, B.~T., {Neugebauer}, G., \& {Houck}, J.~R. 1987, \araa, 25, 187

\bibitem[{{Soifer} {et~al.}(1986){Soifer}, {Sanders}, {Neugebauer},
  {Danielson}, {Lonsdale}, {Madore}, \& {Persson}}]{soi1986}
{Soifer}, B.~T., {Sanders}, D.~B., {Neugebauer}, G., {Danielson}, G.~E.,
  {Lonsdale}, C.~J., {Madore}, B.~F., \& {Persson}, S.~E. 1986, \apjl, 303, L41

\bibitem[{{Stern} {et~al.}(2005){Stern}, {Eisenhardt}, {Gorjian}, {Kochanek},
  {Caldwell}, {Eisenstein}, {Brodwin}, {Brown}, {Cool}, {Dey}, {Green},
  {Jannuzi}, {Murray}, {Pahre}, \& {Willner}}]{ste2005}
{Stern}, D., {et al.} 2005, \apj, 631, 163

\bibitem[{{van Dokkum} {et~al.}(2004){van Dokkum}, {Franx}, {F{\"o}rster
  Schreiber}, {Illingworth}, {Daddi}, {Knudsen}, {Labb{\'e}}, {Moorwood},
  {Rix}, {R{\"o}ttgering}, {Rudnick}, {Trujillo}, {van der Werf}, {van der
  Wel}, {van Starkenburg}, \& {Wuyts}}]{vand2004}
{van Dokkum}, P.~G., {et al.} 2004, \apj, 611, 703

\bibitem[{{van Dokkum} {et~al.}(2006){van Dokkum}, {Quadri}, {Marchesini},
  {Rudnick}, {Franx}, {Gawiser}, {Herrera}, {Wuyts}, {Lira}, {Labb{\'e}},
  {Maza}, {Illingworth}, {F{\"o}rster Schreiber}, {Kriek}, {Rix}, {Taylor},
  {Toft}, {Webb}, \& {Yi}}]{vand2006}
{van Dokkum}, P.~G., {et al.}  2006, \apjl, 638, L59

\bibitem[{{Wake} {et~al.}(2006){Wake}, {Nichol}, {Eisenstein}, {Loveday},
  {Edge}, {Cannon}, {Smail}, {Schneider}, {Scranton}, {Carson}, {Ross},
  {Brunner}, {Colless}, {Couch}, {Croom}, {Driver}, {da {\^A}ngela}, {Jester},
  {de Propris}, {Drinkwater}, {Bland-Hawthorn}, {Pimbblet}, {Roseboom},
  {Shanks}, {Sharp}, \& {Brinkmann}}]{wak2006}
{Wake}, D.~A., {et al.} 2006,
  \mnras, 372, 537

\bibitem[{{Weedman} {et~al.}(2006){Weedman}, {Soifer}, {Hao}, {Higdon},
  {Higdon}, {Houck}, {Le Floc'h}, {Brown}, {Dey}, {Jannuzi}, {Rieke}, {Desai},
  {Bian}, {Thompson}, {Armus}, {Teplitz}, {Eisenhardt}, \& {Willner}}]{wee2006}
{Weedman}, D.~W., {et al.} 2006, \apj,
  651, 101

\bibitem[{{Yan} {et~al.}(2005){Yan}, {Chary}, {Armus}, {Teplitz}, {Helou},
  {Frayer}, {Fadda}, {Surace}, \& {Choi}}]{yan2005}
{Yan}, L., {et al.} 2005, \apj, 628, 604

\bibitem[{{Yan} {et~al.}(2007){Yan}, {Sajina}, {Fadda}, {Choi}, {Armus},
  {Helou}, {Teplitz}, {Frayer}, \& {Surace}}]{yan2007}
{Yan}, L., {et al.} 2007, \apj, 658, 778

\end{thebibliography}

\end{document}